\documentclass[review]{elsarticle}

\usepackage{lineno,hyperref}
\usepackage{caption}
\usepackage{graphicx}
\usepackage[position=t,singlelinecheck=off,skip=-5pt,margin=-5pt]{subcaption}
\usepackage[]{overpic}
\usepackage{pict2e}
\usepackage{amsmath,amssymb}    
\usepackage{times}              
\usepackage{lscape}             
\usepackage{multirow}           
\usepackage{geometry}
\usepackage{float}
\usepackage{hyperref}
\hypersetup{hidelinks,
    colorlinks=true,
    allcolors=black,
    pdfstartview=Fit,
    breaklinks=true}

\modulolinenumbers[1]

\journal{null}

\biboptions{authoryear}

\begin{document}
\captionsetup[figure]{labelfont={bf},
    labelformat={default},
    labelsep=period,
    name={Fig.}}

\begin{frontmatter}

    \title{QIENet: Quantitative irradiance estimation network using recurrent neural network based on satellite remote sensing data}

    \author[myaddress01,myaddress02]{Longfeng Nie}
    \cortext[mycorrespondingauthor]{Corresponding author}
    \author[myaddress03]{Yuntian Chen\corref{mycorrespondingauthor}}
    \ead{ychen@eitech.edu.cn}
    \author[myaddress03,myaddress02,myaddress01]{Dongxiao Zhang\corref{mycorrespondingauthor}}
    \ead{dzhang@eitech.edu.cn}

    \author[myaddress01,myaddress04]{Xinyue Liu}
    \author[myaddress05]{Wentian Yuan}

    \address[myaddress01]{School of Environmental Science and Engineering, Southern University of Science and Technology, Shenzhen 518055, P. R. China}
    \address[myaddress02]{Peng Cheng Laboratory, Shenzhen, 518000, P. R. China}
    \address[myaddress03]{Ningbo Institute of Digital Twin, Eastern Institute of Technology, Ningbo, 315200, P. R. China}
    \address[myaddress04]{Department of Civil and Environmental Engineering, National University of Singapore, Singapore 117576, Singapore}
    \address[myaddress05]{Beijing Kingtansin Technology Company Limited, Beijing, 100000, P. R. China}

    \begin{abstract}
        Global horizontal irradiance (GHI) plays a vital role in estimating solar energy resources, which are used to generate sustainable green energy.
        In order to estimate GHI with high spatial resolution, a quantitative irradiance estimation network, named QIENet, is proposed.
        Specifically, the temporal and spatial characteristics of remote sensing data of the satellite Himawari-8 are extracted and fused by recurrent neural network (RNN) and convolution operation, respectively.
        Not only remote sensing data, but also GHI-related time information (hour, day, and month) and geographical information (altitude, longitude, and latitude), are used as the inputs of QIENet.
        The satellite spectral channels B07 and B11 - B15 and time are recommended as model inputs for QIENet according to the spatial distributions of annual solar energy.
        Meanwhile, QIENet is able to capture the impact of various clouds on hourly GHI estimates.
        More importantly, QIENet does not overestimate ground observations and can also reduce RMSE by 27.51\%/18.00\%, increase $\mathrm{R^{2}}$ by 20.17\%/9.42\%, and increase r by 8.69\%/3.54\% compared with ERA5/NSRDB.
        Furthermore, QIENet is capable of providing a high-fidelity hourly GHI database with spatial resolution $0.02^{\circ}\times0.02^{\circ}$ (approximately $\mathrm{2km \times 2km}$) for many applied energy fields.
    \end{abstract}
\end{frontmatter}

\section{Introduction}
\label{section:1}
Solar energy is a vital energy source for human production and development~\citep{Lord2021598}, especially in the context of rapid and massive consumption of non-renewable energy.
In order to cope with the contradiction between the growing energy demand and policy requirements for energy conservation and emission reduction, photovoltaic power generation has become an important way to generate sustainable green energy~\citep{Kruitwagen2021, Joshi2021}, which not only alleviates a large number of greenhouse gas emissions, but also reduces rural poverty~\citep{Zhang2020}.
Therefore, the distribution of photovoltaic panels needs to be properly optimized according to the distribution of global horizontal irradiance (GHI), and it is of great practical significance to estimate the surface solar irradiance~\citep{HUANG2019111371}.

In order to estimate solar irradiance on the Earth's surface~\citep{JIAO2022102802, OLPENDA2019116}, numerous scientific researches have been conducted in the past, and various models have been designed~\citep{CANO198631, PINKER1995108, BEYER1996207, RIGOLLIER2004159, POLO2012275, GUEYMARD2015379, POLO201625, osti_1778700}, including physical, empirical, semi-empirical, and machine learning methods.
The main purpose of physical models is to establish a relationship between the atmosphere state and solar irradiance based on the radiation transfer model (RTM)~\citep{Dedieu1987, DENEKE20083131, AKITSU2022102724}.
With the assistance of remote sensing satellites, many look-up table (LUT)-based methods~\citep{RONGGAOLIU2008998, ZHANG2014318, HE201520, YU2021102380} were widely applied to calculate solar irradiance.
The LUT-based algorithm involves judgment and calculation of meteorological parameters, such as sky condition, temperature, and aerosol.
However, it is challenging to obtain these parameters in practice, and the physical models are considered to be complex.
In contrast, empirical models are simple and easy to operate, which are generally based on interpolation or regression methods to estimate solar irradiance~\citep{XU2016117, URRACA20171098}.
Semi-empirical models combine some physical attributes on the basis of empirical models to improve accuracy~\citep{CHEN2022404}.
Unlike physical, empirical, and semi-empirical methods, machine learning offers advantage of mining model nonlinearity, and automatically updates the parameters of the model by learning the relationship between atmospheric variables and solar irradiance.

With the development of machine learning, data-driven models are widely used in many disciplines~\citep{Alex6707742, Kumari20211890085, KUMARI2021123285, KUMARI2021117061, LI2022102926, GEORGANOS2022103013, WALDELAND2022102840, LI2023103098, ISHIKAWA2023103215}.
A large number of machine learning methods have been proposed and applied for quantitative irradiance estimation~\citep{HUANG2019111371, KUMARI2021128566, ZHOU2021113960}.
An iterative random forest model was established using high spatiotemporal resolution satellite data to estimate and map half-hour solar radiation with 1-km spatial resolution in the continental United States~\citep{CHEN2021916}.
Based on satellite imagery, an artificial neural network (ANN) was applied to elucidate the nonlinear physical relationship between remote sensing data and daily GHI observations~\citep{LU20113179, QUESADARUIZ2015494}.
Not only remote sensing data, but also meteorological factors~\citep{QUEJ201762}, can be used as inputs into machine learning models.
\citet{LINARESRODRIGUEZ20115356} and \citet{WANG2016384} used meteorological factors to predict solar radiation values by ANN.
\citet{MARZO2017303} took daily extraterrestrial radiation and minimum and maximum temperatures as inputs of an ANN model to estimate daily GHI in the desert.
Meanwhile, \citet{MIRA2016251} obtained surface net radiation ($\mathrm{R_{n}}$) through the input surface variables, including emissivity, albedo, and temperature derived from Landsat-7 imagery, and evaluated the uncertainty of $\mathrm{R_{n}}$.
Moreover, there are some radiation estimation studies that have combined intelligent algorithms and machine learning.
\citet{OLATOMIWA2015632} proposed an SVM-FFA model based on support vector machine (SVM) and the firefly algorithm (FFA) to estimate monthly mean GHI accurately.
The genetic algorithm (GA)-ANN model was developed to estimate all-sky daily average surface all-wave $\mathrm{R_{n}}$ at high latitudes from remote sensing data~\citep{WANG201531, CHEN2020111842}.
Considering the spatial characteristics of remote sensing data~\citep{JIANG2019109327, JIANG2020115178, Yeom_2020, Jang14081840, Xu20222315, Kumaresan2022v1}, the convolutional neural network (CNN) was adopted to estimate or predict solar irradiance.

Although machine learning has made great progress in solar irradiance research, certain problems remain to be solved.
The models, which are trained through meteorological data monitored by sparse ground stations, cannot estimate GHI in the area without ground stations.
Since remote sensing data offer the advantages of wide coverage, multi-spectrum, and high spatiotemporal resolution, they are recommended to use in this study.
More importantly, the GHI observations of the ground-based pyranometers are accurate, but sparse, in both spatial and temporal scales, which are typical multi-source high-fidelity data.
Therefore, there is no way to directly obtain the GHI distribution of the whole region.

Considering the aforementioned problems, the main purpose for this paper is to develop a quantitative irradiance estimation network, named QIENet, using standard/convolutional recurrent neural network (RNN), and establish a high-fidelity spatiotemporal GHI database based on remote sensing data of the satellite Himawari-8 in the study area. In section~\ref{section:Datasets}, the study area and data processing are introduced in detail. Section~\ref{section:Methodology} shows the architecture of QIENet and quantitative evaluation metrics of the current model, and the results of QIENet are discussed in section~\ref{section:Results_Discussion}. Finally, some important conclusions are summarized.

\section{Datasets}
\label{section:Datasets}
\subsection{Ground-based hourly GHI observations} 
\label{datasets:Irradiance}
\begin{figure}[H]
    \centering
    \vspace{0mm}
    \subcaptionbox{\label{fig:study_region:a}}{
        \includegraphics[width=0.53\textwidth,trim=0 0 0 0,clip]{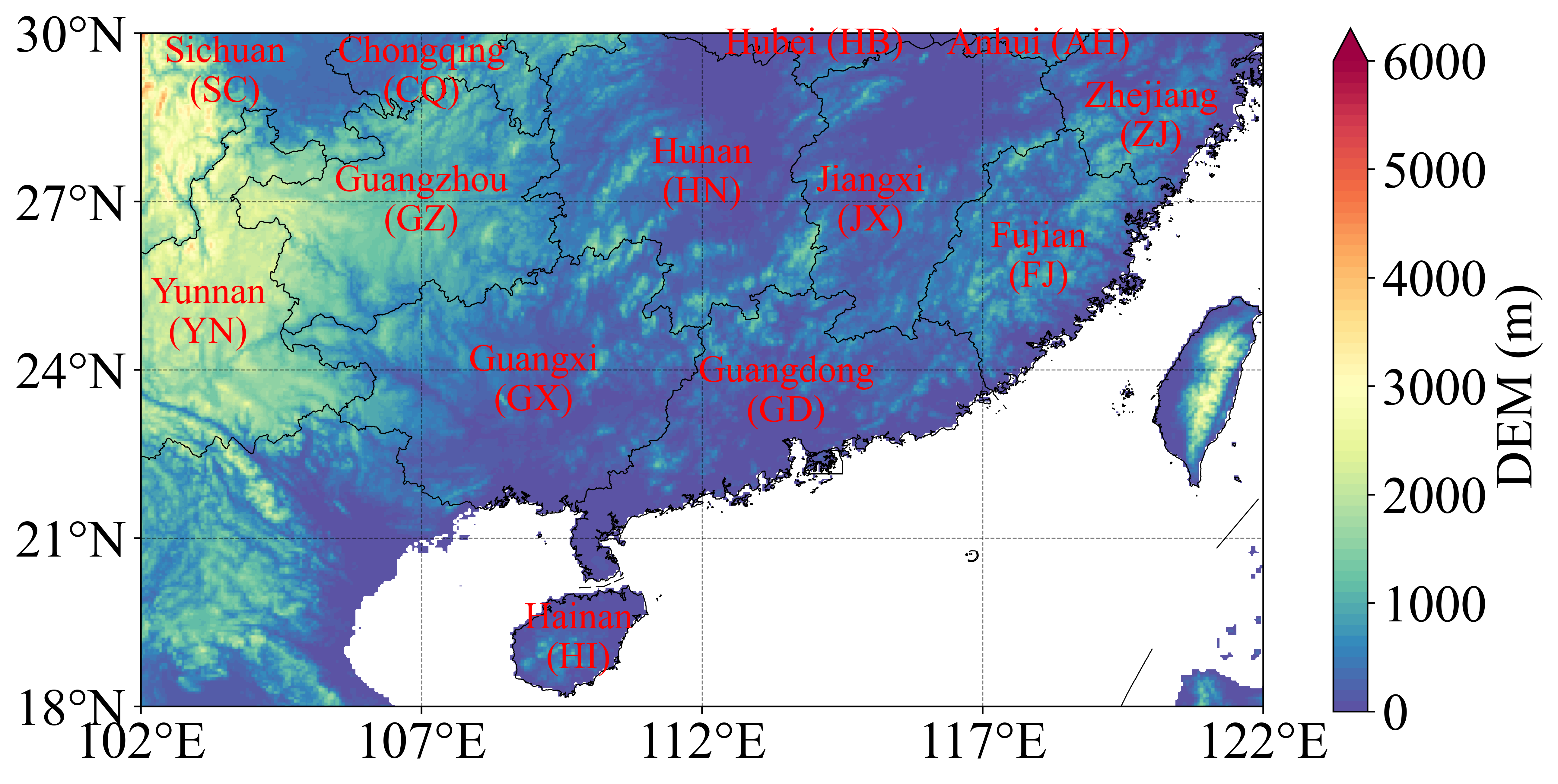}
    }
    \hfill
    \subcaptionbox{\label{fig:study_region:b}}{
        \includegraphics[width=0.4\textwidth,trim=0 0 0 0,clip]{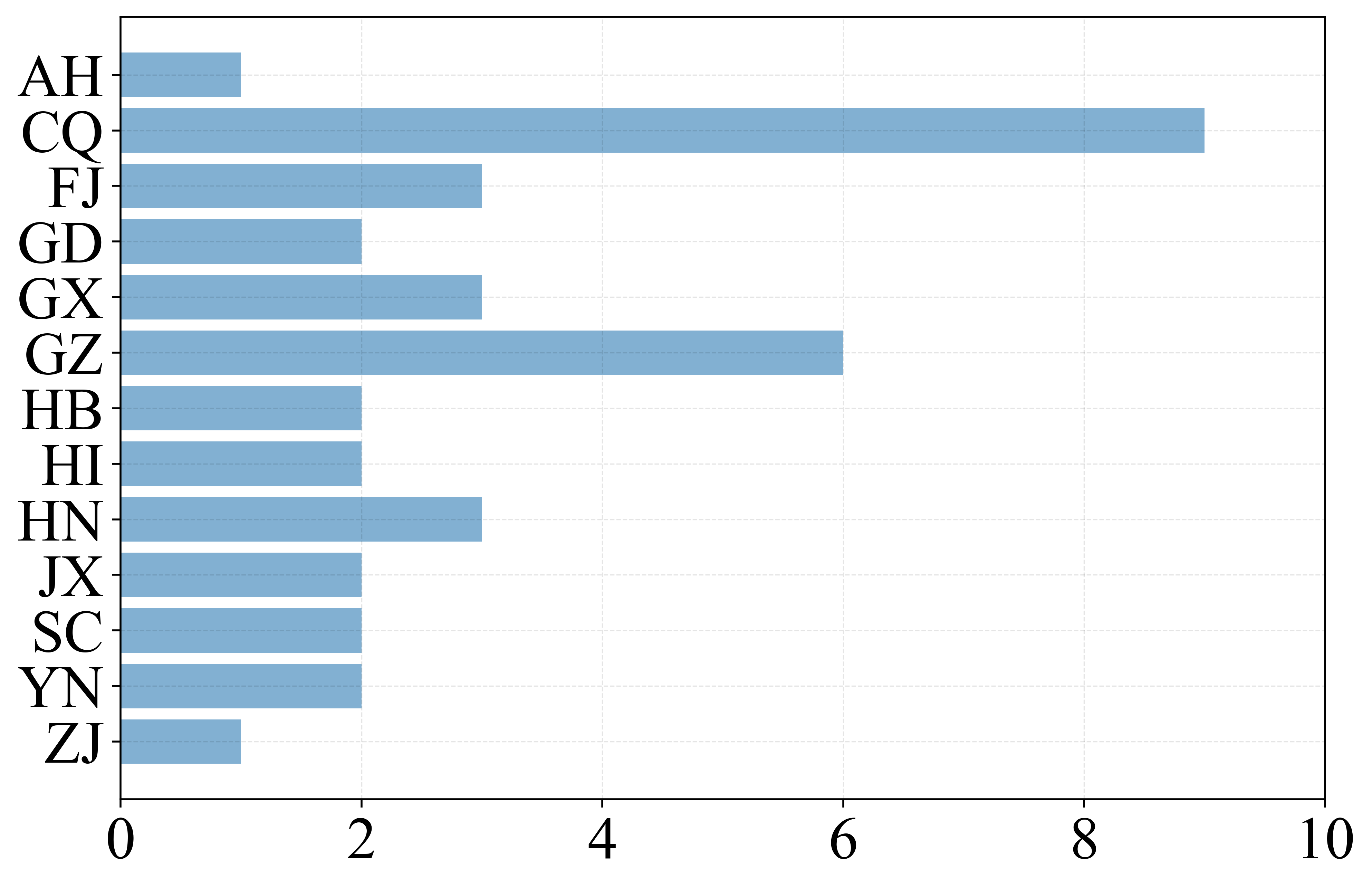}
    }
    \caption{(a) The scope of the study area, and (b) the quantitative statistics of 38 observation stations in southern provinces of China.}
    \label{fig:study_region}
\end{figure}

There are 38 GHI observation stations in the study area (102° - 122° E, 18° - 30° N), shown in Fig.~\ref{fig:study_region:a}, which provide hourly ground measurements of the Earth's surface solar irradiance.
These stations cover 13 provinces in southern China.
Fig.~\ref{fig:study_region:b} shows the number of monitoring stations in each province~\citep{FAN2019168, Zhang202010}.
Missing values often exist because of recording equipment failure.
Here, a total of 639,999 hourly GHI observations are recorded from July 1, 2019 to June 30, 2021,
and the hourly GHI observation is an hourly average of instantaneous monitoring values of the pyranometers, which have a sampling frequency of 1 min and are maintained once a month, including calibration and cleaning.
Concurrently, the quality of hourly GHI values is guaranteed through two steps, including the physical threshold method~\citep{MORADI20091, TANG2010466} and outlier detection.

\begin{figure}[H]
    \centering
    \vspace{-0mm}
    \includegraphics[width=.70\textwidth]{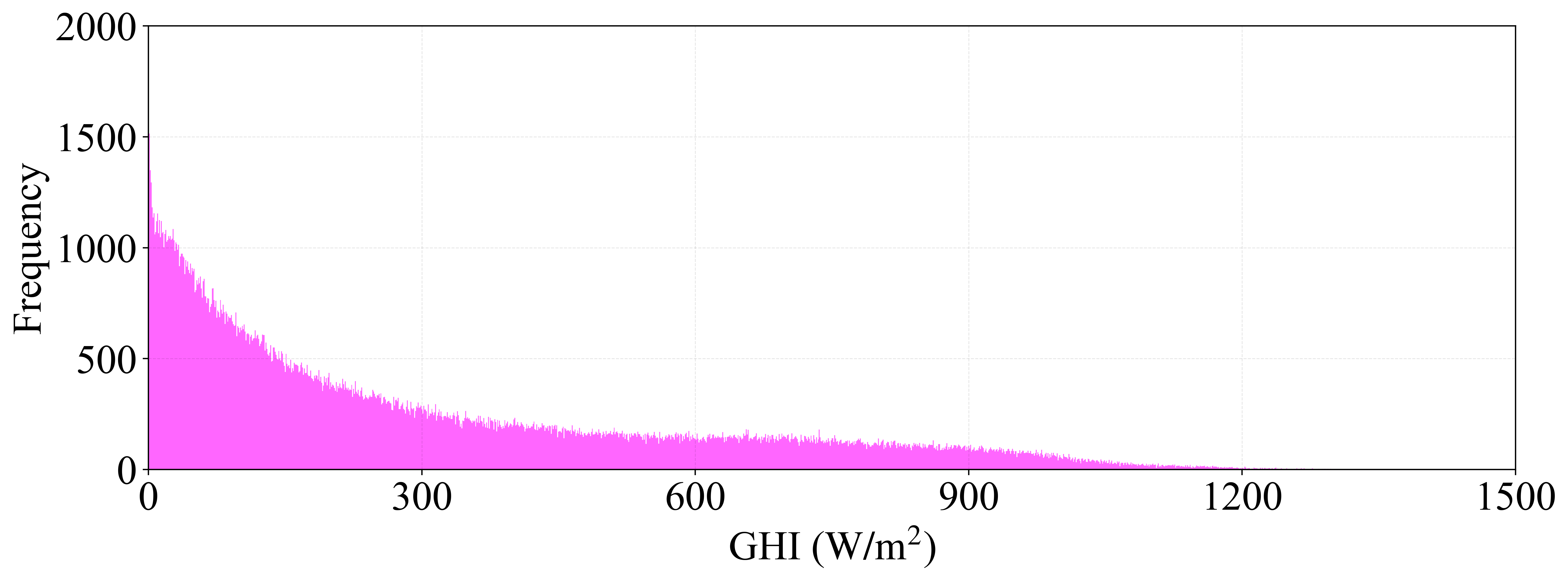}
    \caption{Histograms of frequency distributions for valid hourly GHI observations.}
    \label{fig:irrad_freq}
\end{figure}

The physical threshold method implements the requirement that GHI observations shall not exceed extraterrestrial GHI, and 95.89\% (613,664/639,999) of hourly GHI observations meet this requirement.
Outlier detection makes use of interquartile range (IQR) to remove outliers of hourly GHI observations at the same hour, which are larger than the upper whisker.
The upper whisker is calculated by $\mathrm{Q_3+1.5\times(Q_3-Q_1)}$, where $\mathrm{Q_1}$ and $\mathrm{Q_3}$ are the lower and upper quartiles of GHI observations at the same hour, respectively.
In total, 605,928 hourly GHI observations are retained, comprising 94.68\% (605,928/639,999) of the original GHI observations.
Histograms of frequency distributions of these valid GHI observations are presented in Fig.~\ref{fig:irrad_freq}.

\subsection{Satellite remote sensing data}
\label{datasets:Satellite}
The satellite Himawari-8, operated by the Japan Meteorological Agency, was launched on October 7, 2014, and whose end-of-life is not less than 2030. It is the first flight unit of the Himawari 3rd generation programme and replaced the Multi-functional Transport Satellite-2 as an operational satellite on July 7, 2015.
The main mission of Himawari-8 is operational meteorology, and its additional mission is environmental applications~\citep{ZHANG2023103506, HU2020102153, ZHAO2022102933}.
Details of spectral channels of the satellite Himawari-8 are introduced in Table~\ref{table:AHI_info}.
In this research, Himawari L1 gridded data within the study area were downloaded through the Japan Aerospace Exploration Agency (JAXA) Himawari Monitor P-Tree System (available at \href{https://www.eorc.jaxa.jp/ptree/index.html}{https://www.eorc.jaxa.jp/ptree/index.html}).
The spatial resolution is $0.02^{\circ}\times0.02^{\circ}$ (approximately $\mathrm{2km \times 2km}$), and the temporal resolution is 10 min.
Satellite spectral channels include albedo of B01 - B06 and brightness temperature (BT) of B07 - B16.

\begin{figure}[!htbp]
    \centering
    \vspace{-0mm}
    \includegraphics[width=.70\textwidth]{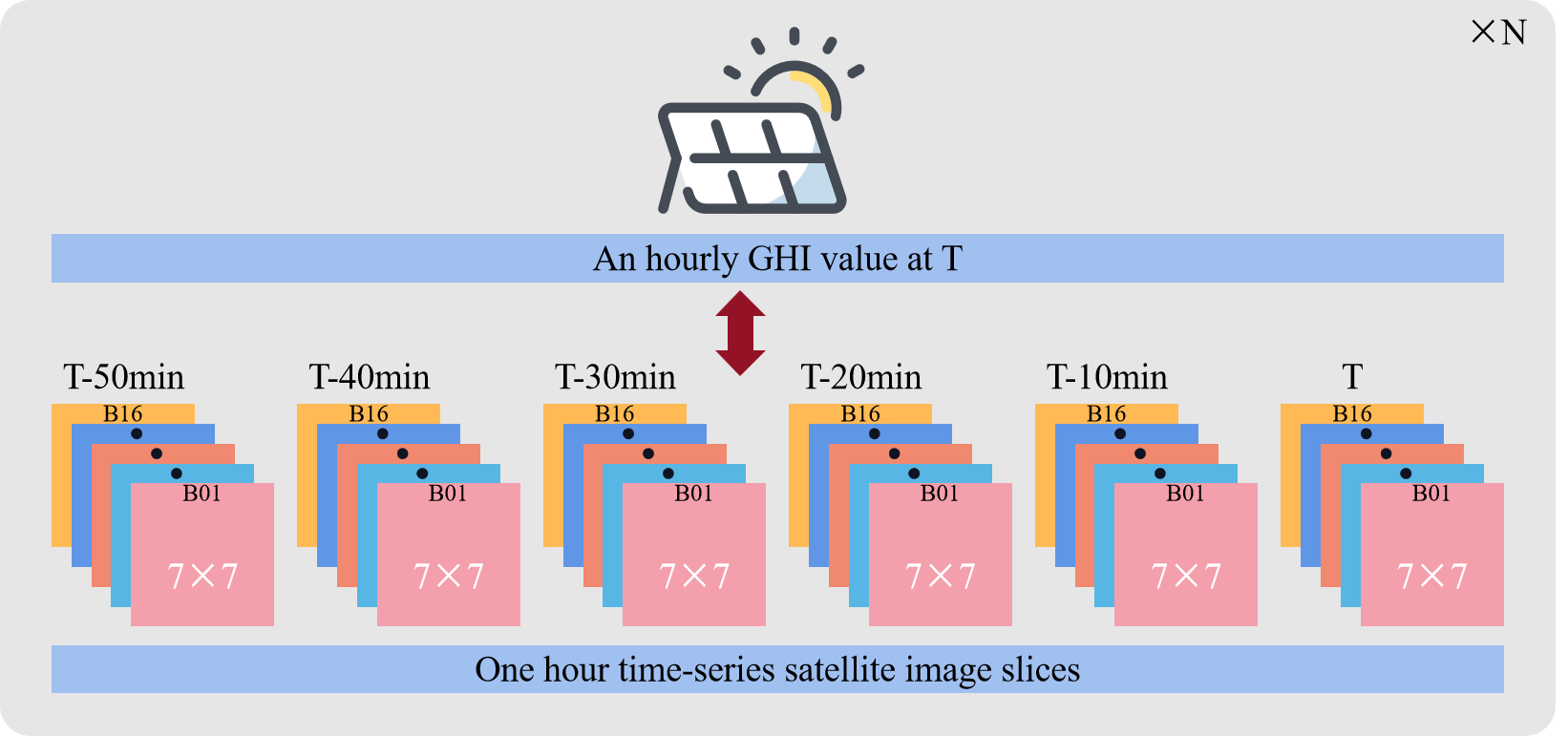}
    \caption{Illustration of an hourly GHI value corresponding to one hour time-series satellite image slices.}
    \label{fig:GHI_RS}
\end{figure}

For an hourly GHI estimate, the $\mathrm{7\times7}$ pixel-sized satellite image slices are required, as shown in Fig.~\ref{fig:GHI_RS}.
This is because clouds 6 km from the monitoring station move over the station after 10 min, thus affecting the sunlight~\citep{Hakuba2014JD021946, JIANG2019109327}.
A one-to-one dataset of satellite image slices and valid GHI observations was constructed.
The matrix shapes for satellite image slices and hourly GHI observations are N~$\times$~T~$\times$~C~$\times$~H~$\times$~W (598639~$\times$~6~$\times$~16~$\times$~7~$\times$~7) and N~$\times$~O (598639~$\times$~1), respectively, where N is the sample size of the dataset, T is the number of time-series satellite image slices in 1 h, C is the number of satellite spectral channels, H and W are the height and width of the satellite image slice, respectively, and O is the dimension of hourly GHI.
The relationship distributions among hourly GHI observations and the maximum, minimum, mean, and median values for each channel of the satellite image slices are depicted in Fig.~\ref{fig:irrad_Albedo_TBB}.
From Fig.~\ref{fig:irrad_Albedo_TBB}a and~\ref{fig:irrad_Albedo_TBB}c, some abnormal distributions are mainly caused by problematic satellite remote sensing data (as shown in Fig.~\ref{fig:satellite_problem}).
After discarding the faulty satellite remote sensing data, 576,122 samples were retained, accounting for 96.24\% (576,122/598,639).
The relationship distributions among hourly GHI observations and the maximum, minimum, mean, and median values for each channel of the satellite image slices become reasonable, as shown in Fig.~\ref{fig:irrad_Albedo_TBB}b and~\ref{fig:irrad_Albedo_TBB}d.

\section{Methodology}
\label{section:Methodology}
\subsection{Quantitative irradiance estimation network} 
\label{sub:QIENet model}
\begin{figure}[!hbp]
    \centering
    \includegraphics[width=0.96\textwidth,trim=0 0 0 0,clip]{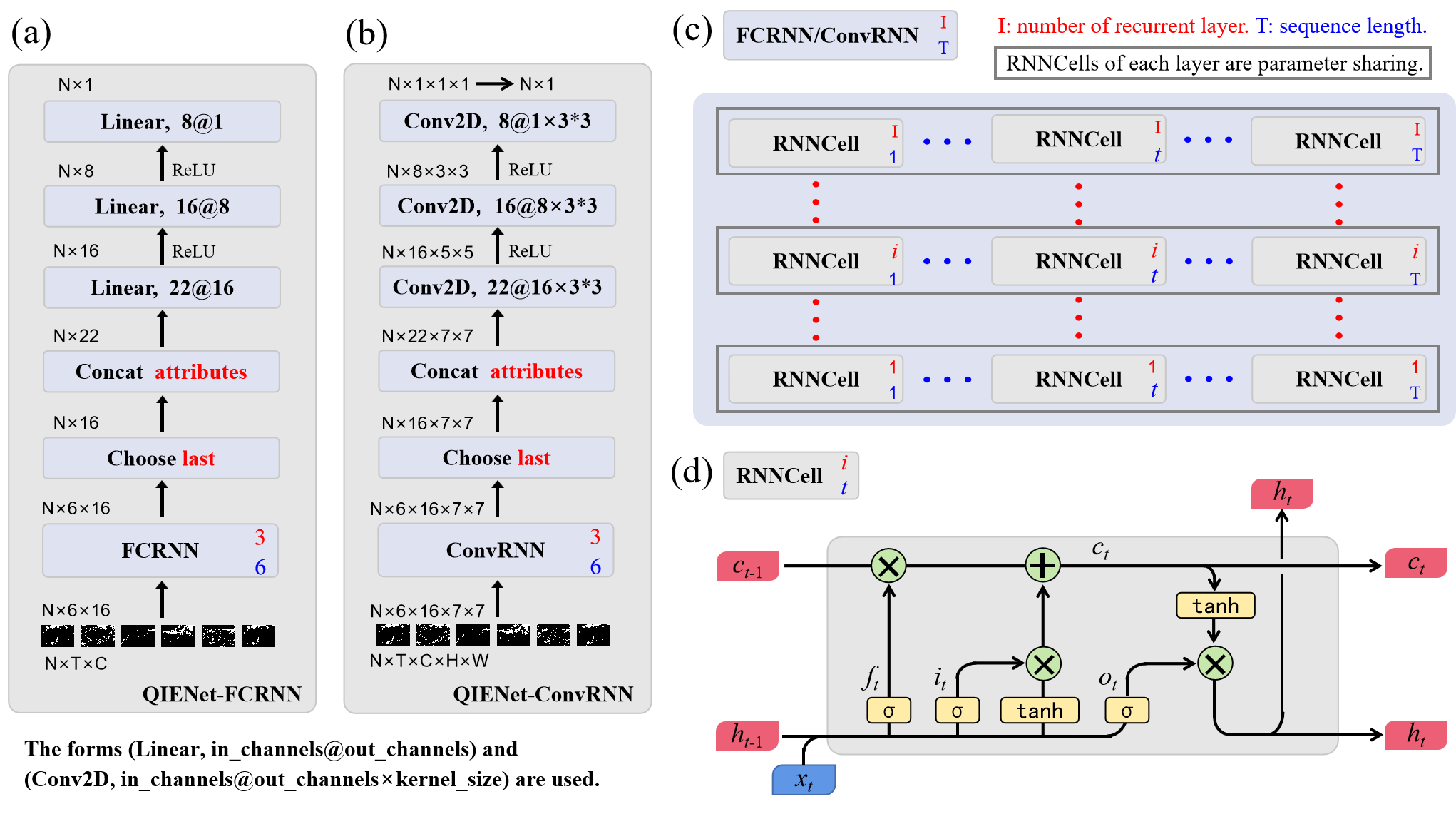}
    \caption{The architecture of the RNN-based quantitative irradiance estimation network, named QIENet, including (a) QIENet-FCRNN, (b) QIENet-ConvRNN, (c) multilayer RNN, and (d) LSTM-based RNNCell.}
    \label{fig:models}
\end{figure}

In order to estimate hourly GHI with high spatial resolution $0.02^{\circ}\times0.02^{\circ}$, a quantitative irradiance estimation network, named QIENet, is proposed and applied in the study area.
The framework of the proposed QIENet is illustrated in Fig.~\ref{fig:models}, which contains two network forms: QIENet-FCRNN using the fully-connected RNN in Fig.~\ref{fig:models}a, and QIENet-ConvRNN using the convolutional RNN in Fig.~\ref{fig:models}b.
Compared with QIENet-FCRNN, QIENet-ConvRNN not only considers the temporal features of remote sensing data, but also integrates spatial features.
In Fig.~\ref{fig:models}a and~\ref{fig:models}b, `last' refers to the last time sequence output of the multilayer RNN,
`attributes' includes GHI-related information, such as time (hour, day, and month) and geographical information (altitude, longitude, and latitude),
and the activation function `ReLU' is used to enhance the nonlinear expression ability of the model.
The architecture of the multilayer RNN is shown in Fig.~\ref{fig:models}c, whose single unit, named RNNCell, using the widely adopted LSTM, is depicted in Fig.~\ref{fig:models}d.

At present, RNN has two operation modes~\citep{Graves2013LSTM, shi2015convLSTM}: fully connect (FCRNN) and convolution (ConvRNN). FCRNN is mentioned in the work of \citet{Graves2013LSTM}, and the specific expressions are equations (\ref{eq:f_t}) - (\ref{eq:h_t}), respectively:

\begin{equation}
    \label{eq:f_t}
    f_{t} = \sigma (W_{xf} x_{t} + W_{hf} h_{t-1} + b_{f})
\end{equation}
\begin{equation}
    \label{eq:i_t}
    i_{t} = \sigma (W_{xi} x_{t} + W_{hi} h_{t-1} + b_{i})
\end{equation}
\begin{equation}
    \label{eq:c_t}
    c_{t} = f_{t} \circ c_{t-1} + i_{t} \circ \tanh (W_{xc} x_{t} + W_{hc} h_{t-1} + b_{c})
\end{equation}
\begin{equation}
    \label{eq:o_t}
    o_{t} = \sigma (W_{xo} x_{t} + W_{ho} h_{t-1} + b_{o})
\end{equation}
\begin{equation}
    \label{eq:h_t}
    h_{t} = o_{t} \circ \tanh(c_{t})
\end{equation}
where $f_{t}$, $i_{t}$, and $o_{t}$ represent the forget gate, input gate, and output gate, respectively; $h_{t}$ is the hidden state; $c_{t}$ is the memory cell; $W$ and $b$ are weight and bias, respectively; the symbol $\circ$ denotes Hadamard product; and $x$ is the input. \citet{shi2015convLSTM} replaced the fully connected operation with the convolution operation to adapt the input of image-type data, whose specific forms are as follows:

\begin{equation}
    \label{eq:conv:f_t}
    f_{t} = \sigma (W_{xf} * x_{t} + W_{hf} * h_{t-1} + b_{f})
\end{equation}
\begin{equation}
    \label{eq:conv:i_t}
    i_{t} = \sigma (W_{xi} * x_{t} + W_{hi} * h_{t-1} + b_{i})
\end{equation}
\begin{equation}
    \label{eq:conv:c_t}
    c_{t} = f_{t} \circ c_{t-1} + i_{t} \circ \tanh (W_{xc} * x_{t} + W_{hc} * h_{t-1} + b_{c})
\end{equation}
\begin{equation}
    \label{eq:conv:o_t}
    o_{t} = \sigma (W_{xo} * x_{t} + W_{ho} * h_{t-1} + b_{o})
\end{equation}
\begin{equation}
    \label{eq:conv:h_t}
    h_{t} = o_{t} \circ \tanh(c_{t})
\end{equation}

Equations (\ref{eq:conv:f_t}) - (\ref{eq:conv:h_t}) are different from equations (\ref{eq:f_t}) - (\ref{eq:h_t}), in that * represents the convolution operator. There are two forms of the multilayer RNN, including $\mathrm{FCRNN_{T}^{I}}$ and $\mathrm{ConvRNN_{T}^{I}}$, where T is the length of the time-series, and I is the number of layers of the multilayer RNN. Compared with QIENet-FCRNN, QIENet-ConvRNN is used to determine whether remote sensing spatial fusion can improve the accuracy of our model.

The constructed dataset in section~\ref{section:Datasets} is divided into two parts, including the dataset from July 2019 to June 2020 and the dataset from July 2020 to June 2021.
The dataset from July 2019 to June 2020 is divided into training and validation sets by using the five-fold cross-validation method, which makes the evaluated QIENet models more accurate and credible for training.
At the same time, an overfitting problem often occurs during the model training process for deep learning~\citep{Dietterich1995, OH2022122921}; therefore, the model training process takes advantage of early stopping to avoid overfitting.
The process of early stopping is that the training of QIENet is conducted on the training set, and mean square error (MSE) of the model on the validation set is calculated every other epoch.
The specific criterion of early stopping is that the MSE of the validation dataset no longer decreases after 15 consecutive epochs.
The dataset from July 2020 to June 2021 is used to evaluate the performance of QIENet.

\subsection{Quantitative evaluation metrics} 
\label{sub:Quantitative evaluation metrics}
In order to quantitatively evaluate the performance of QIENet in estimating GHI, RMSE, $\mathrm{R^2}$, mean bias error (MBE), and correlation coefficient (r) are selected as quantitative error indicators.
Among them, r with the range of [-1, +1] is used to measure the correlation between the estimated GHIs and the observed GHIs, and $\mathrm{R^2}$ assesses the degree of agreement between them.
For MBE, a negative value implies underestimation, and vice versa.
Therefore, the larger are the $\mathrm{R^2}$ and r, the better is the model; the smaller is the RMSE, the better is the model; and the closer the MBE is to zero, the better is the model.
The best possible results are RMSE = 0, MBE = 0, $\mathrm{R^{2}}$ = 1, and r = 1.
These indicators are defined by the following equations (\ref{eq:RMSE}) - (\ref{eq:r}), respectively:

\begin{equation}
    \label{eq:RMSE}
    \mathrm{RMSE}=\sqrt{\frac{\sum_{i=1}^{\mathrm{n}}\left(Ir_{\mathrm{est}}^{i}-Ir_{\mathrm{obs}}^{i}\right)^{2}}{\mathrm{n}}}
\end{equation}
\begin{equation}
    \label{eq:MBE}
    \mathrm{MBE}=\frac{1}{\mathrm{n}}{\sum_{i=1}^{n}\left(Ir_{\mathrm{est}}^{i}-Ir_{\mathrm{obs}}^{i}\right)}
\end{equation}
\begin{equation}
    \label{eq:R2}
    \mathrm{R^{2} }=1-\frac{\sum_{i=1}^{\mathrm{n}}\left(Ir_{\mathrm{obs}}^{i}-Ir_{\mathrm{est}}^{i}\right)^{2} }{\sum_{i=1}^{\mathrm{n}}\left(Ir_{\mathrm{obs}}^{i}-Ir_{\mathrm{obs}}^{-}\right)^{2} }
\end{equation}
\begin{equation}
    \label{eq:r}
    \mathrm{r}=\frac{\sum_{i=1}^{\mathrm{n}}\left(Ir_{\mathrm{est}}^{i}-Ir_{\mathrm{est}}^{-}\right)\left(Ir_{\mathrm{obs}}^{i}-Ir_{\mathrm{obs}}^{-}\right)}{\sqrt{\sum_{i=1}^{\mathrm{n}}\left(Ir_{\mathrm{est}}^{i}-Ir_{\mathrm{est}}^{-}\right)^{2} \sum_{i=1}^{\mathrm{n}}\left(Ir_{\mathrm{obs}}^{i}-Ir_{\mathrm{obs}}^{-}\right)^{2}}}
\end{equation}
where $Ir_{\mathrm {obs }}^{i}$ denotes GHI measurement from pyranometers located in the study area; $Ir_{\mathrm {est }}^{i}$ denotes the estimated value of the current model; $Ir_{\mathrm {obs }}^{-}$ and $Ir_{\mathrm {est }}^{-}$ represent the mean of hourly GHI observations and estimates, respectively; and $\mathrm{n}$ represents the number of hourly GHI samples.


\section{Results and discussion}
\label{section:Results_Discussion}
\subsection{The performance of QIENet in estimating GHI} 
\label{sub:Performance of QIENet}

\begin{figure}[!htbp]
    \vskip-0pt
    \centering
    \subcaptionbox{\vspace{0mm}\label{fig:QIENet_Date_geography:M1012035}}{
        \includegraphics[width=0.90\textwidth,trim=0 0 0 0,clip]{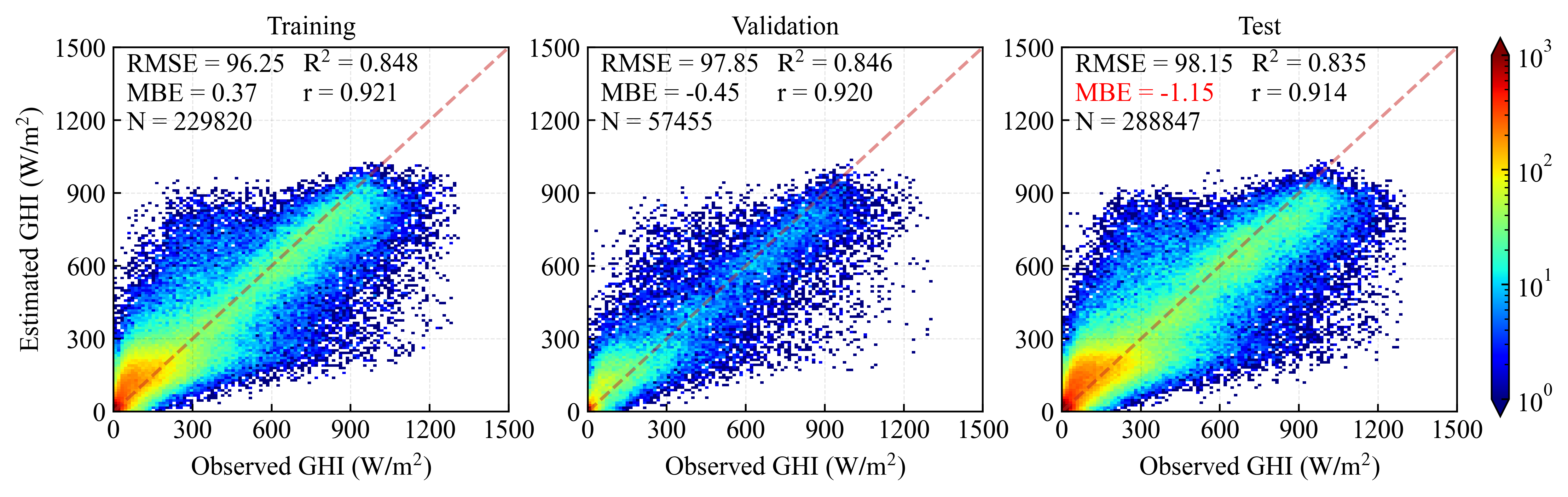}
    }
    \subcaptionbox{\vspace{0mm}\label{fig:QIENet_Date_geography:M1022035}}{
        \includegraphics[width=0.90\textwidth,trim=0 0 0 0,clip]{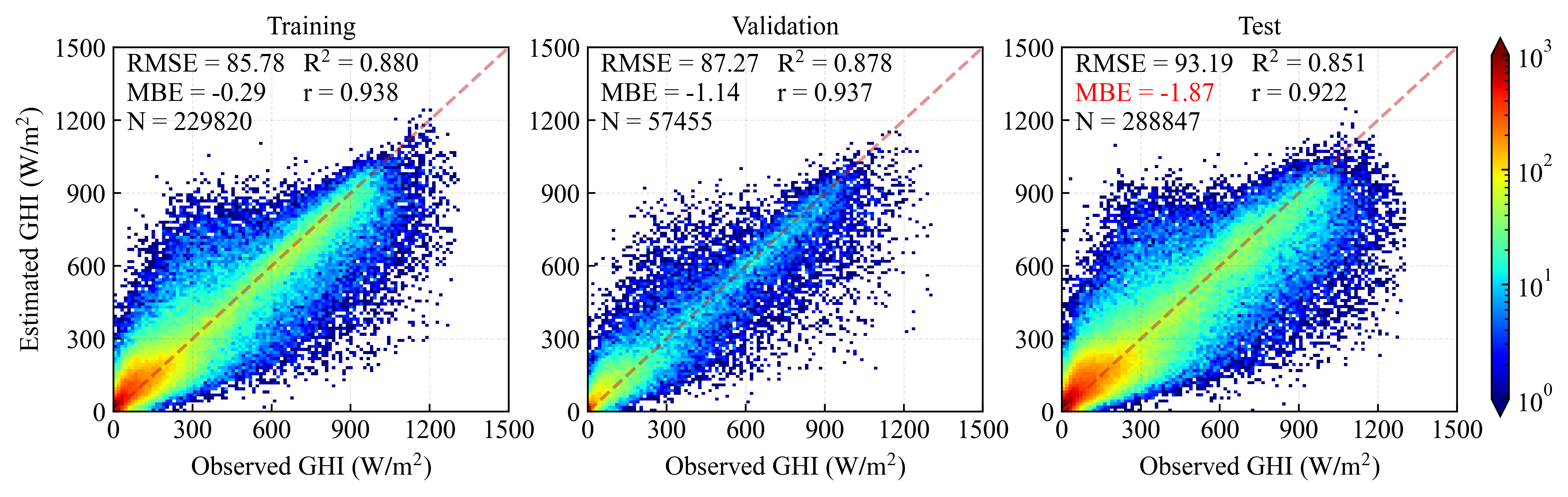}
    }
    \subcaptionbox{\vspace{0mm}\label{fig:QIENet_Date_geography:JAXA}}{
        \includegraphics[width=0.90\textwidth,trim=0 0 0 0,clip]{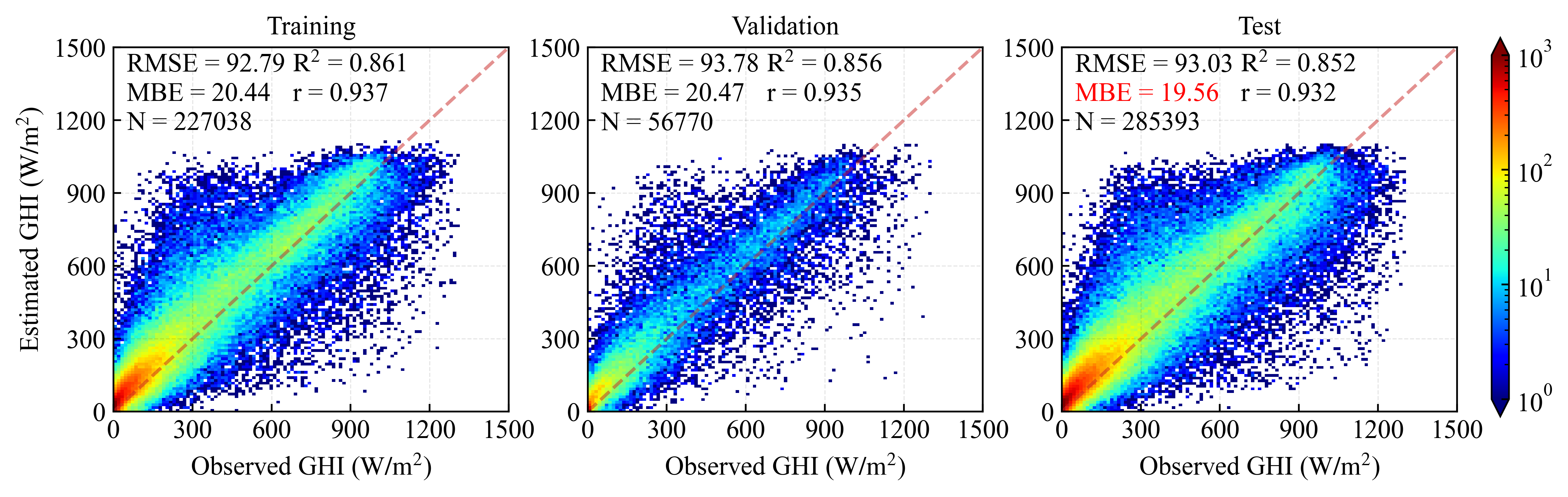}
    }
    \subcaptionbox{\vspace{0mm}\label{fig:QIENet_Date_geography:ERA5}}{
        \includegraphics[width=0.90\textwidth,trim=0 0 0 0,clip]{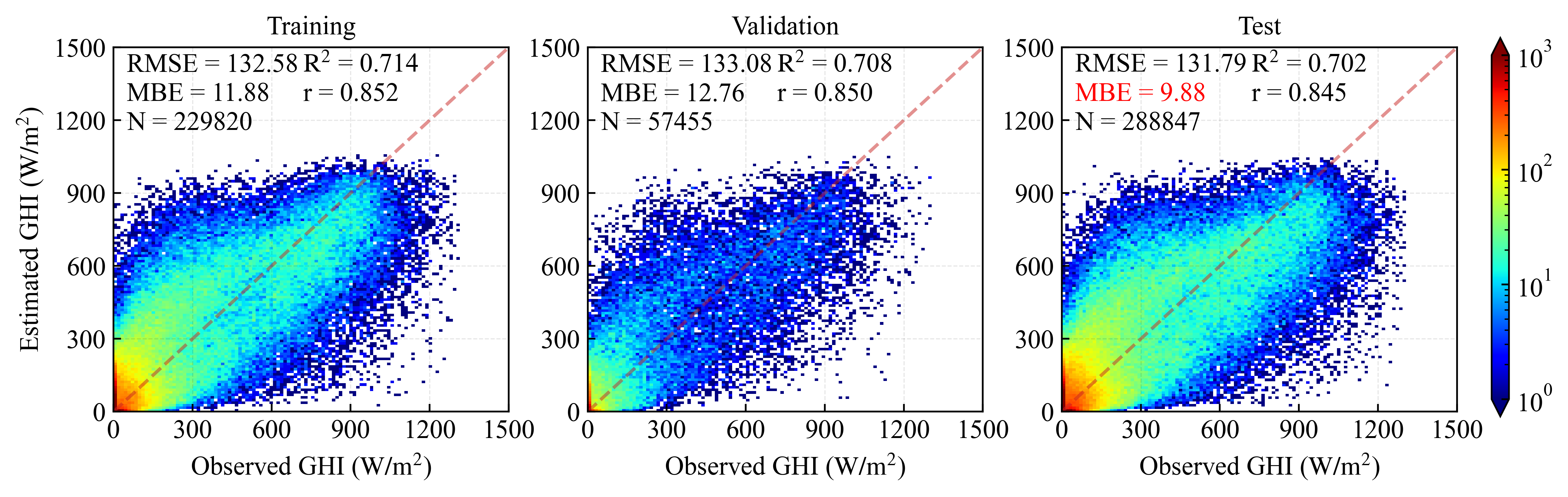}
    }
    \caption{2D histograms of ground-based hourly GHI observations to estimated GHI values using (a) QIENet\_FC1, (b) QIENet\_Conv1, (c) the JAXA product, (d) ERA5, and (e) NSRDB. The number of bins in both dimensions is 100 and the color bar denotes the number of points contained in each bin. The red dotted-line is the one-to-one reference line.}
    \label{fig:QIENet_Date_geography}
\end{figure}
\begin{figure}[!htbp]
    \vspace{-5mm}
    \centering
    \ContinuedFloat    
    \subcaptionbox{\vspace{0mm}\label{fig:QIENet_Date_geography:NSRDB}}{
        \includegraphics[width=0.90\textwidth,trim=0 0 0 0,clip]{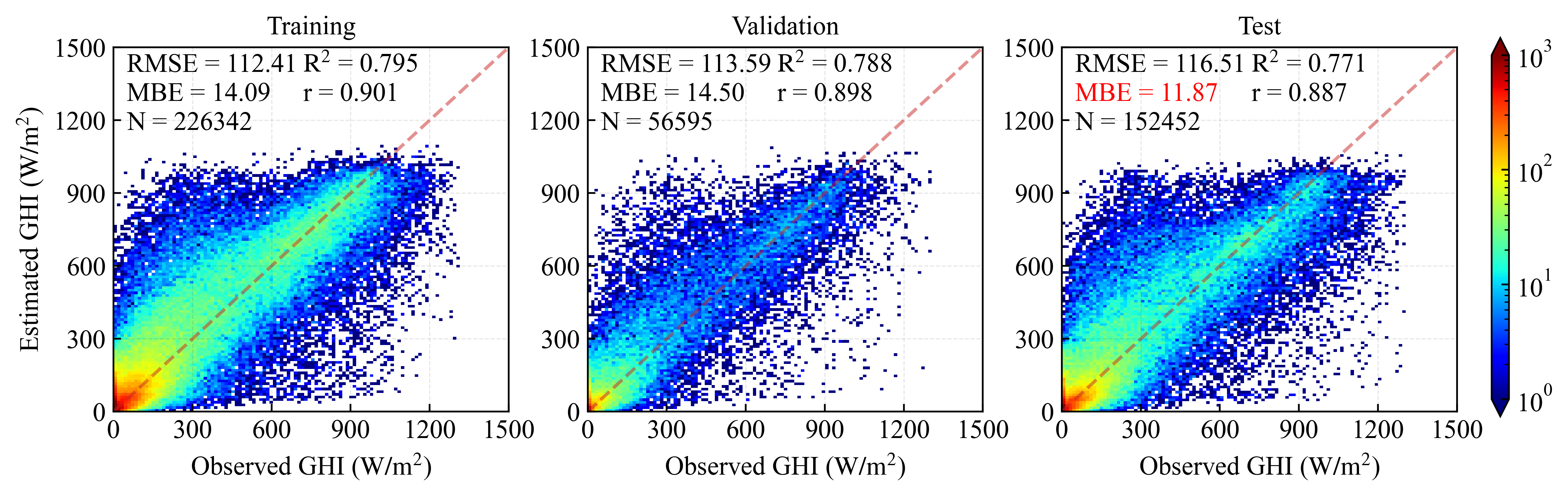}
    }
    \caption{Continued}
\end{figure}
\setcounter{subfigure}{0}

According to network type and the input variables of the model, different QIENet models are presented in Table~\ref{table:model_inputs}.
In this section, the QIENet\_FC1 and QIENet\_Conv1, which use all satellite spectral channels, time, and geography as inputs, are discussed in detail as examples.
The training, validation, and test results of QIENet\_FC1 and QIENet\_Conv1 are shown in Fig.~\ref{fig:QIENet_Date_geography:M1012035} and~\ref{fig:QIENet_Date_geography:M1022035}, respectively.
The model parameters are updated during the training process,
the early stopping technique is utilized to prevent QIENet\_FC1 and QIENet\_Conv1 from overfitting on the validation set, and the performance of the models in estimating GHI is evaluated in the test stage.
Meanwhile, the JAXA product about short wave radiation~\citep{Damiani2018, Takenaka2011} from \href{https://www.eorc.jaxa.jp/ptree/index.html}{the JAXA Himawari Monitor P-Tree System}, the widely used public dataset ERA5 containing hourly GHIs from the European Centre for Medium-Range Weather Forecasts (\href{https://cds.climate.copernicus.eu}{ECMWF}), and the National Solar Radiation Database (NSRDB)~\citep{SENGUPTA201851} using the Physical Solar Model (PSM) of National Renewable Energy Laboratory (NREL) are used to compare with ground-based hourly GHI observations during training, validation, and test, respectively, as shown in Fig.~\ref{fig:QIENet_Date_geography:JAXA}, \ref{fig:QIENet_Date_geography:ERA5}, and~\ref{fig:QIENet_Date_geography:NSRDB}.
The evaluation indicators on the test dataset are recorded in Table~\ref{table:QIENet_SIndex}.

\begin{figure}[!htbp]
    \centering
    \subcaptionbox{}{
        \includegraphics[width=0.8\textwidth,trim=0 0 0 0,clip]{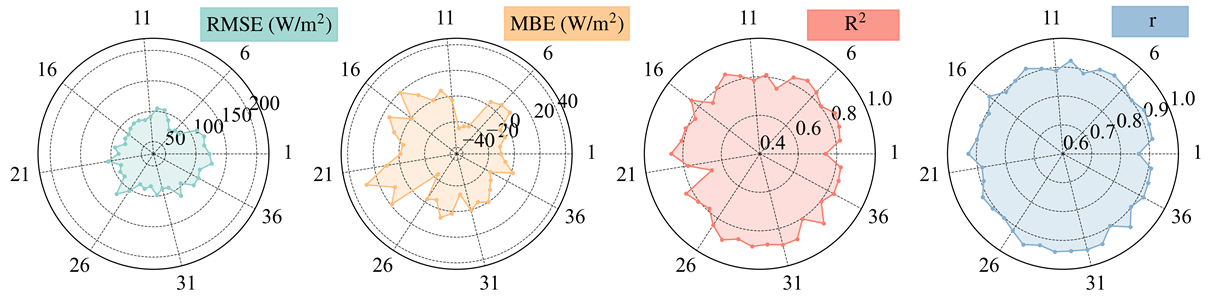}
    }
    \subcaptionbox{}{
        \includegraphics[width=0.8\textwidth,trim=0 0 0 0,clip]{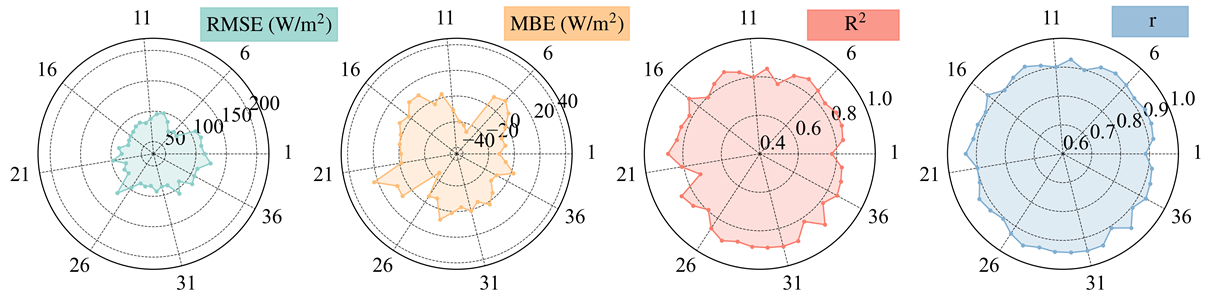}
    }
    \subcaptionbox{}{
        \includegraphics[width=0.8\textwidth,trim=0 0 0 0,clip]{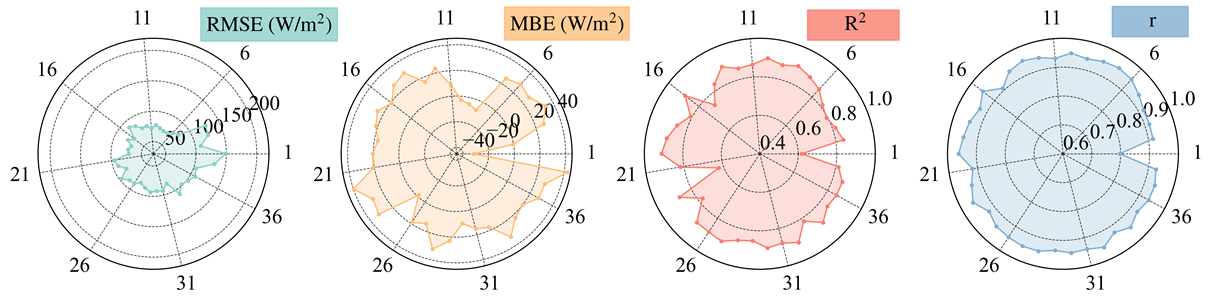}
    }
    \subcaptionbox{}{
        \includegraphics[width=0.8\textwidth,trim=0 0 0 0,clip]{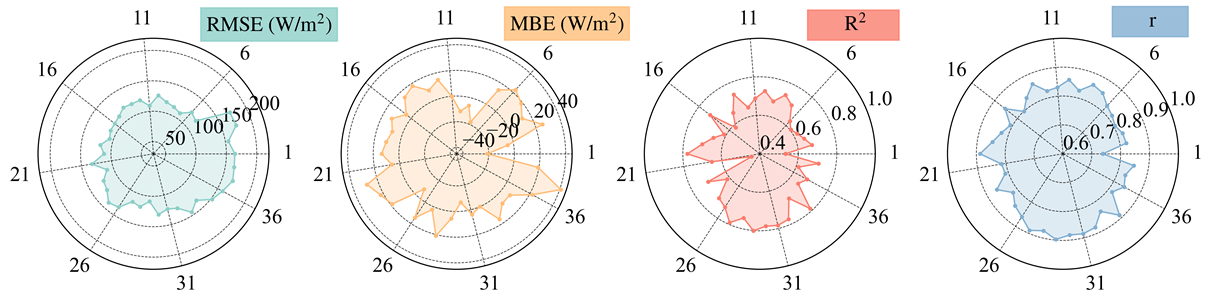}
    }
    \subcaptionbox{}{
        \includegraphics[width=0.8\textwidth,trim=0 0 0 0,clip]{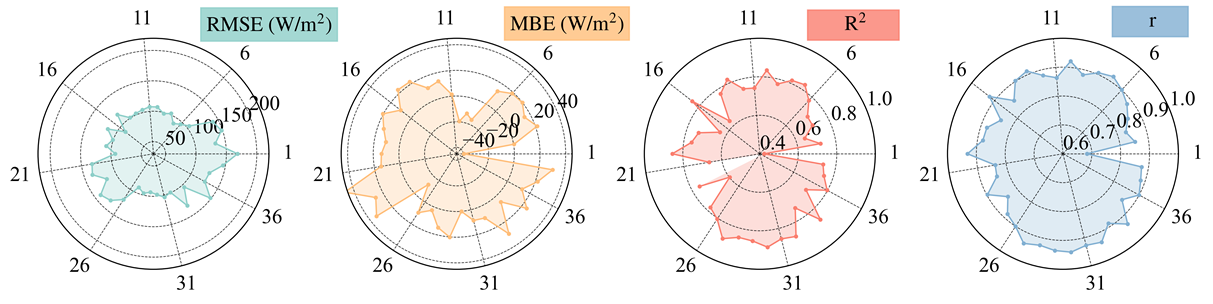}
    }
    \caption{Radar distribution diagrams of evaluation indicators at each station using (a) QIENet\_FC1, (b) QIENet\_Conv1, (c) the JAXA product, (d) ERA5, and (e) NSRDB.}
    \label{fig:index_GHI_station}
\end{figure}

As shown in Fig.~\ref{fig:QIENet_Date_geography},
QIENet\_FC1 and QIENet\_Conv1 are able to reduce RMSE by 25.50\% and 29.17\%, increase $\mathrm{R^{2}}$ by 18.91\% and 21.18\%, and increase r by 8.12\% and 9.15\%, respectively, compared with ERA5 in the test stage.
Meanwhile, QIENet\_FC1 and QIENet\_Conv1 can also reduce RMSE by 15.73\% and 19.88\%, increase $\mathrm{R^{2}}$ by 8.28\% and 10.34\%, and increase r by 3.00\% and 3.97\%, respectively, compared with NSRDB in the test stage.
QIENet\_Conv1 outperforms QIENet\_FC1, which indicates that the spatial fusion of remote sensing information is effective.
In Fig.~\ref{fig:QIENet_Date_geography:M1022035}, the proposed model QIENet\_Conv1 achieves RMSE = 93.19 $\mathrm{W/m^{2}}$, $\mathrm{R^{2}}$ of 0.851 and MBE of -1.87 $\mathrm{W/m^{2}}$. Compared with the JAXA product (RMSE = 93.03 $\mathrm{W/m^{2}}$, $\mathrm{R^{2}}$ = 0.852 and MBE = 19.56 $\mathrm{W/m^{2}}$), the RMSE of QIENet\_Conv1 only decreases by 0.16 $\mathrm{W/m^{2}}$ (0.171\%), and the $\mathrm{R^{2}}$ of QIENet\_Conv1 only decreases by 0.001 (0.133\%).
However, it should be noticed that the MBE of the JAXA product is more than 10 times larger than that of QIENet\_Conv1, indicating that the predictions of QIENet are more unbiased.
The fact that the JAXA product is often overestimated has been pointed out in the work of \citet{Damiani2018}.
Our model QIENet does not overestimate ground observations, which has a smaller estimation bias for ground observations compared with the JAXA product, ERA5, and NSRDB.
Due to the crucial importance of unbiased predictions in practical applications, and considering the one-order-of-magnitude improvement in MBE achieved by QIENet, it can be concluded that QIENet exhibits better performance.
Therefore, QIENet is a good choice in estimating hourly GHI.

Radar distribution diagrams of evaluation indicators at each station using QIENet\_FC1, QIENet\_Conv1, the JAXA product, ERA5, and NSRDB are shown in Fig.~\ref{fig:index_GHI_station}, which are summarized in Table~\ref{table:QIENet_SIndex_everystation}.
The number of ground-based GHI observation stations corresponds to the number of points in the radar diagram.
The average RMSE values of all stations using QIENet\_FC1, QIENet\_Conv1, the JAXA product, ERA5, and NSRDB are 97.00±13.50 $\mathrm{W/m^{2}}$, 91.96±13.55 $\mathrm{W/m^{2}}$, 90.67±18.73 $\mathrm{W/m^{2}}$, 129.83±19.77 $\mathrm{W/m^{2}}$, and 115.12±19.95 $\mathrm{W/m^{2}}$, respectively.
Compared with ERA5, the average RMSE values of QIENet\_FC1, QIENet\_Conv1, the JAXA product, and NSRDB are reduced by 25.28\%, 29.17\%, 30.16\%, and 11.33\%, respectively.
The average $\mathrm{R^{2}}$ values of all stations using QIENet\_FC1, QIENet\_Conv1, the JAXA product, ERA5, and NSRDB are 0.822±0.046, 0.841±0.037, 0.844±0.064, 0.684±0.075, and 0.751±0.111, respectively.
Compared with ERA5, the average $\mathrm{R^{2}}$ values of QIENet\_FC1, QIENet\_Conv1, the JAXA product, and NSRDB are increased by 20.24\%, 23.03\%, 23.35\%, and 9.88\%, respectively.
The average $\mathrm{r}$ values of all stations using QIENet\_FC1, QIENet\_Conv1, the JAXA product, ERA5, and NSRDB are 0.912±0.019, 0.921±0.015, 0.935±0.026, 0.841±0.032, and 0.885±0.046, respectively.
Compared with ERA5, the average $\mathrm{r}$ values of QIENet\_FC1, QIENet\_Conv1, the JAXA product, and NSRDB are increased by 8.42\%, 9.47\%, 11.08\%, and 5.24\%, respectively.
In terms of RMSE, $\mathrm{R^{2}}$ and r, QIENet\_FC1, QIENet\_Conv1, and the JAXA product achieved excellent performance at each station.
For MBE, however, the performance of QIENet is best compared with the JAXA product, ERA5 and NSRDB, and the JAXA product, ERA5 and NSRDB overestimate ground observations.

\subsection{The effects of satellite spectral channels, time, and geographical information}
\label{sub:the_effects_of_input_variables}

\begin{figure}[!htbp]
    \centering
    \subcaptionbox{\label{fig:QIENet_Date_geography:RMSE}}{
        \includegraphics[width=0.80\textwidth,trim=0 0 0 0,clip]{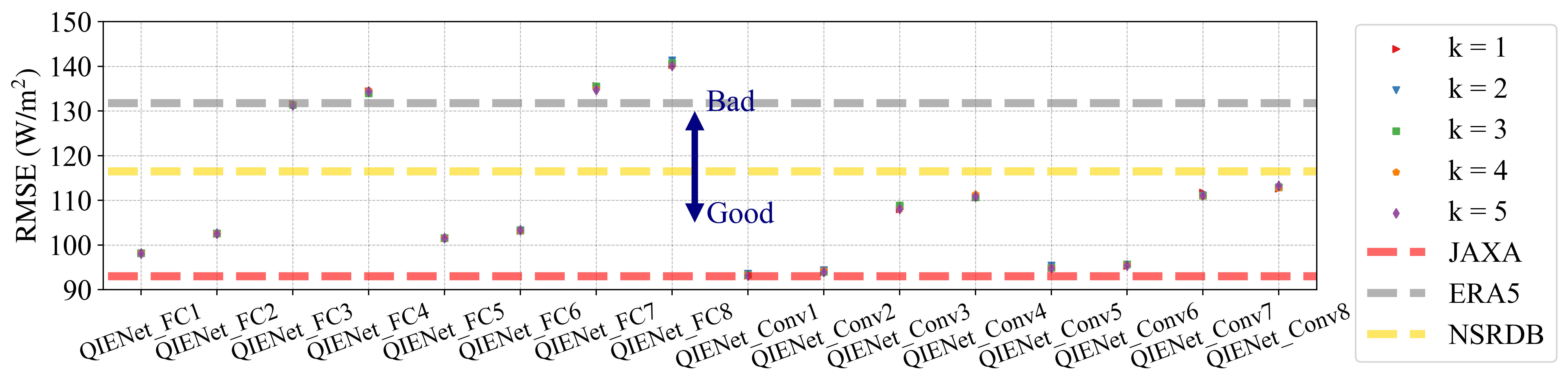}
    }
    \subcaptionbox{\label{fig:QIENet_Date_geography:MBE}}{
        \includegraphics[width=0.80\textwidth,trim=0 0 0 0,clip]{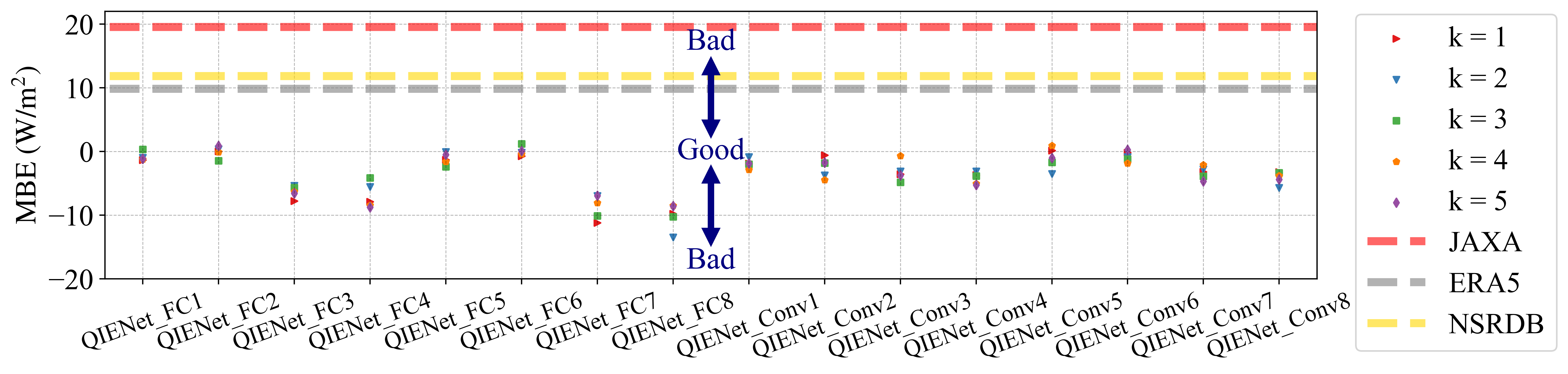}
    }
    \subcaptionbox{\label{fig:QIENet_Date_geography:R2}}{
        \includegraphics[width=0.80\textwidth,trim=0 0 0 0,clip]{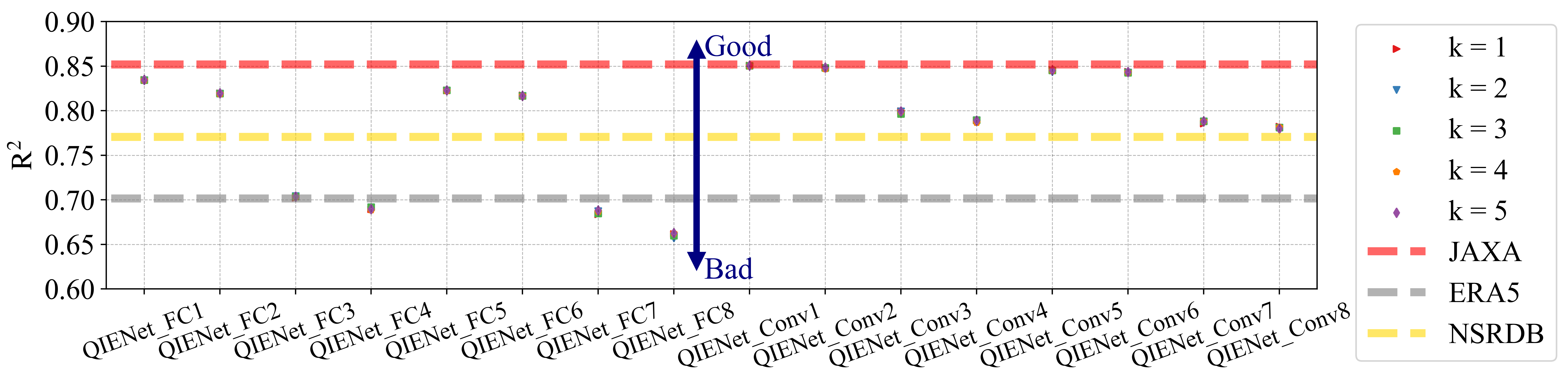}
    }
    \subcaptionbox{\label{fig:QIENet_Date_geography:r}}{
        \includegraphics[width=0.80\textwidth,trim=0 0 0 0,clip]{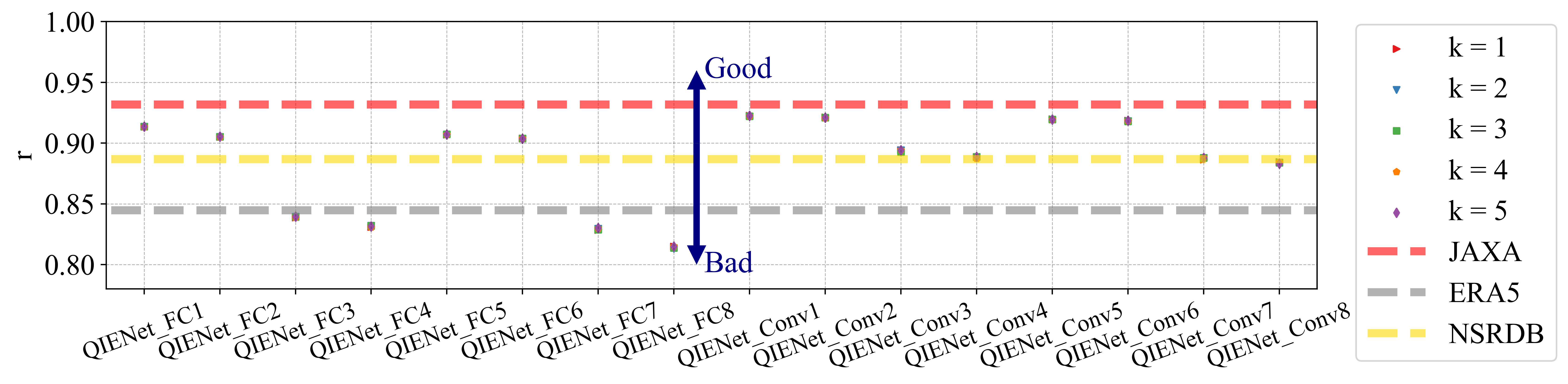}
    }
    \caption{The influence of satellite spectral channels (albedo of B01 - B06 and BT of B07 - B16), time (hour, day, and month), and geographical information (altitude, longitude, and latitude) on the performances of QIENet-FCRNN and QIENet-ConvRNN in terms of (a) RMSE, (b) MBE, (c) $\mathrm{R^2}$, and (d) $\mathrm{r}$. $\mathrm{k}$ is the order of cross validation.}
    \label{fig:FCRNN_ConvRNN_Test_Index}
\end{figure}

To explore the effects of satellite spectral channels (albedo of B01 - B06 and BT of B07 - B16), time (hour, day, and month), and geographical information (altitude, longitude, and latitude) on QIENet, various models in Table~\ref{table:model_inputs} are designed.
Pearson correlation coefficients (PCCs) between the hourly GHI observations and remote sensing data of each satellite spectral channel are calculated, which are listed in Table~\ref{table:band_pearson}.
According to the result of PCCs, BT of B07 and B11 - B15 are selected as model inputs.
\citet{TANA2023113548} made use of infrared spectral information to obtain cloud microphysical properties, and then the GHI is estimated with cloud microphysical properties.
This indicates that BT of B07 and B11 - B15, belonging to the infrared spectrum channels, are effective for estimating GHI.
Fig.~\ref{fig:FCRNN_ConvRNN_Test_Index} illustrates the influence of satellite spectral channels, time, and geographical information on the performances of QIENet-FCRNN and QIENet-ConvRNN.
The evaluation indicators of five-fold cross-validation for each model show that the architectures of QIENet are stable and reliable.

The influence trend of input variables on QIENet-FCRNN and QIENet-ConvRNN is similar in Fig.~\ref{fig:FCRNN_ConvRNN_Test_Index}.
From the results of all designed QIENet models using different input variables, the performance of QIENet using all satellite spectral channels, time, and geographical information is best, and the performance of QIENet only using remote sensing data is worst.
Under the same input conditions, QIENet-ConvRNN outperforms QIENet-FCRNN in estimating GHI, especially for the models not using the time variable.
Compared with the models using all satellite spectral channels, the accuracy of the models using satellite spectral channels B07 and B11 - B15 exhibits just a very slight drop, so that some satellite channels can be overlooked for simplicity of the model.
The results of models QIENet\_FC2, QIENet\_FC6, QIENet\_Conv2, and QIENet\_Conv6 imply that the removal of geographical information has little impact on QIENet because $\mathrm{R^{2}}$ and r decline slightly.
However, the results of models QIENet\_FC3, QIENet\_FC7, QIENet\_Conv3, and QIENet\_Conv7 decreased significantly compared with those of models QIENet\_FC1, QIENet\_FC5, QIENet\_Conv1, and QIENet\_Conv5 when the input time variable is not used.
Essentially, all satellite spectral channels can be replaced by B07 and B11 - B15, and time is more important as an input variable into QIENet than geographical information. Furthermore, using the convolution operation is more effective in improving model accuracy compared to the fully connected operation.

\subsection{Overall solar energy distributions at different time scales}
\label{sub:SolarEnergyDistribution}
\begin{figure}[!htbp]
    \centering
    \subcaptionbox{\label{fig:irrad_map_year:M1012035}}{
        \includegraphics[width=0.31\textwidth,trim=0 0 0 0,clip]{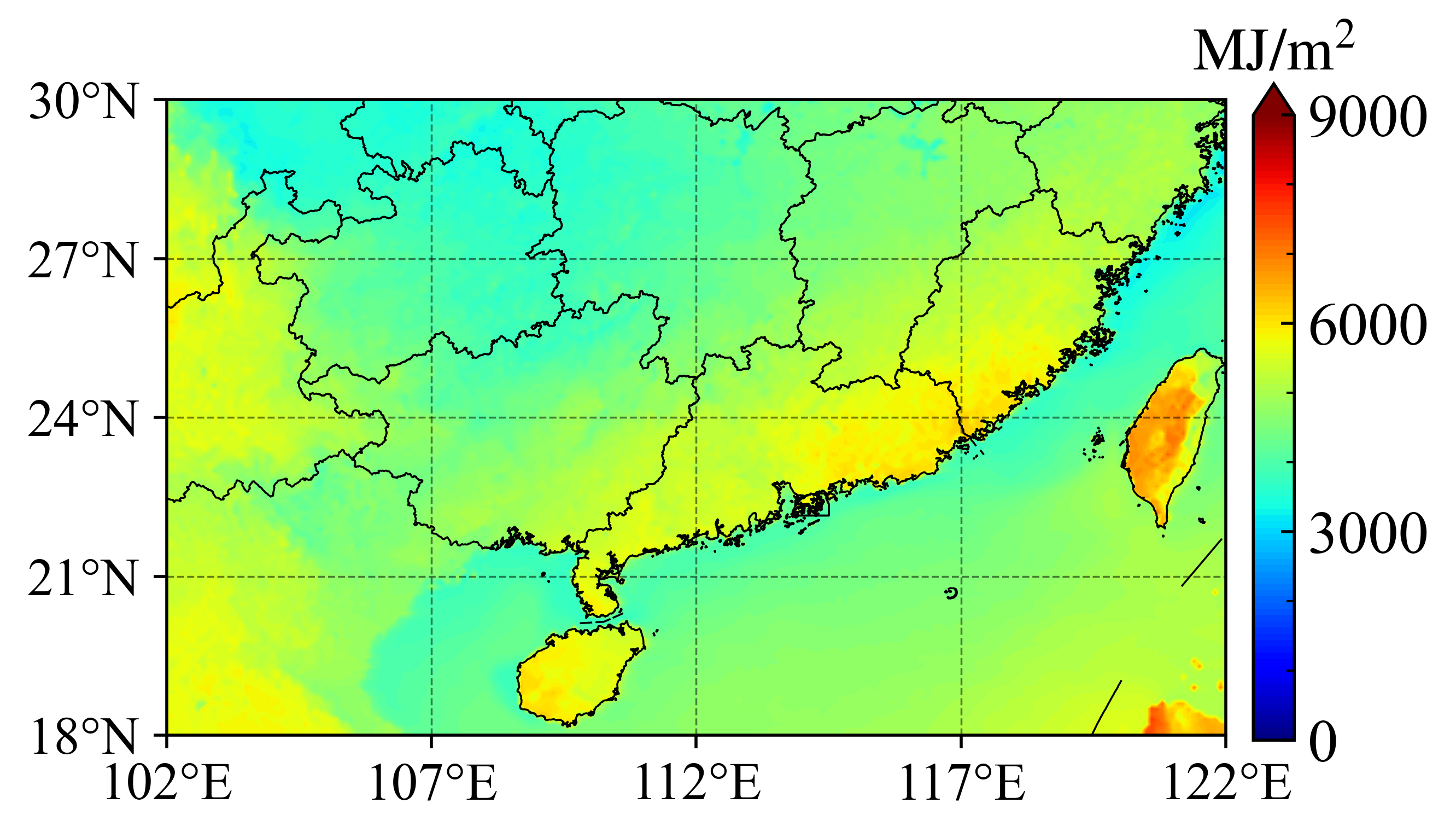}
    }
    \subcaptionbox{\label{fig:irrad_map_year:M1012036}}{
        \includegraphics[width=0.31\textwidth,trim=0 0 0 0,clip]{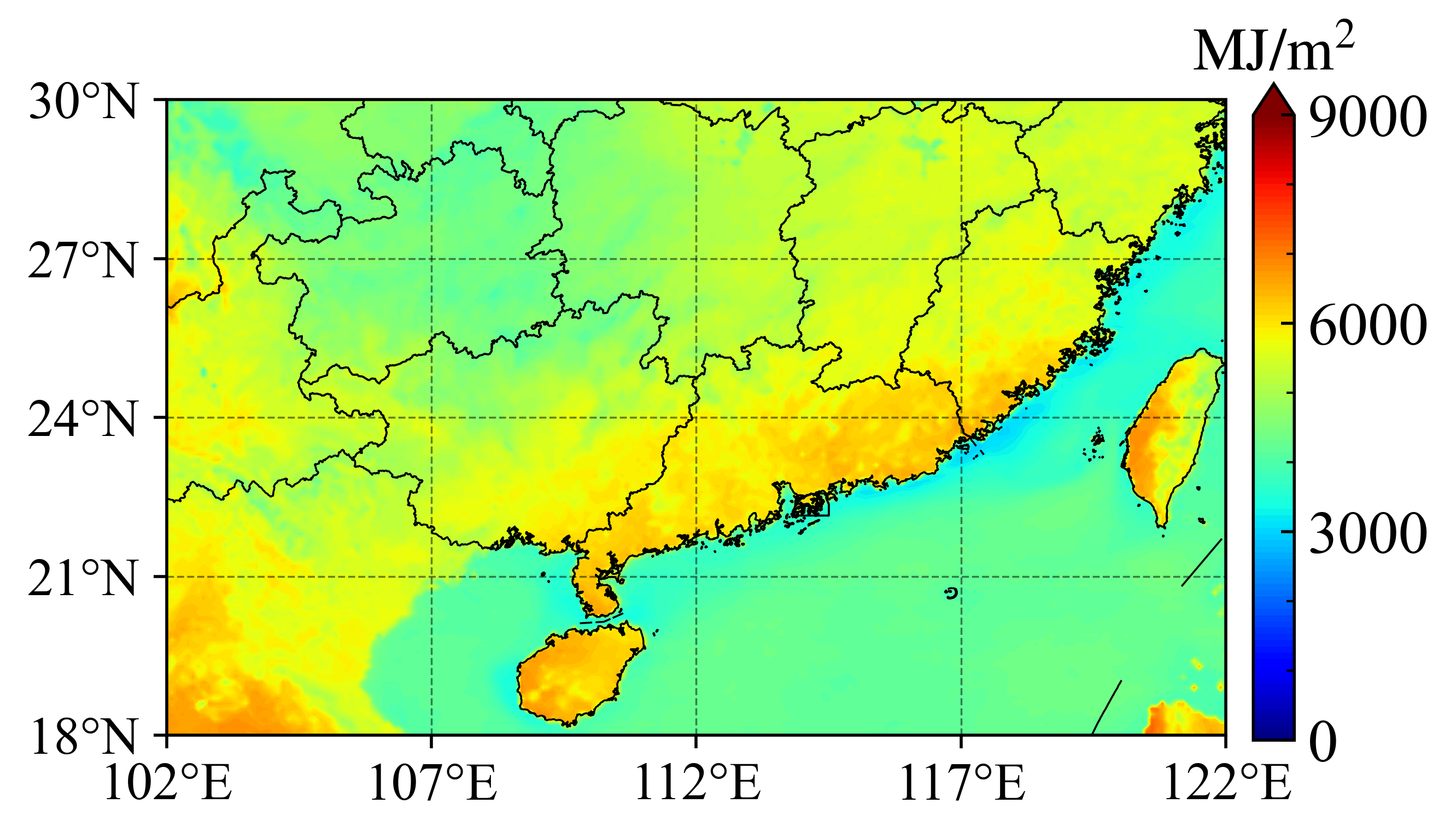}
    }
    \subcaptionbox{\label{fig:irrad_map_year:M1012037}}{
        \includegraphics[width=0.31\textwidth,trim=0 0 0 0,clip]{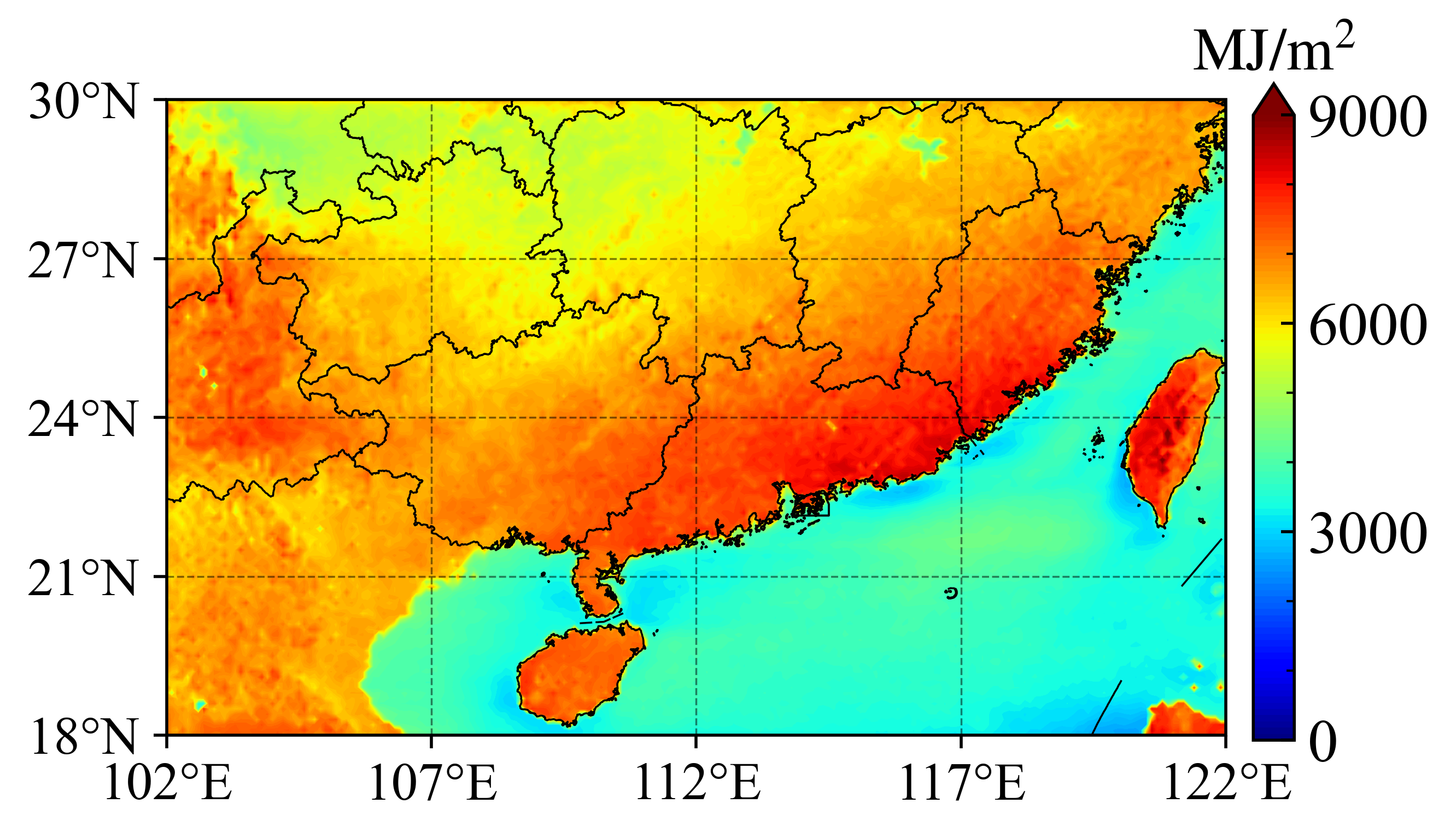}
    }
    \vspace{-1mm}
    \subcaptionbox{\label{fig:irrad_map_year:M1012038}}{
        \includegraphics[width=0.31\textwidth,trim=0 0 0 0,clip]{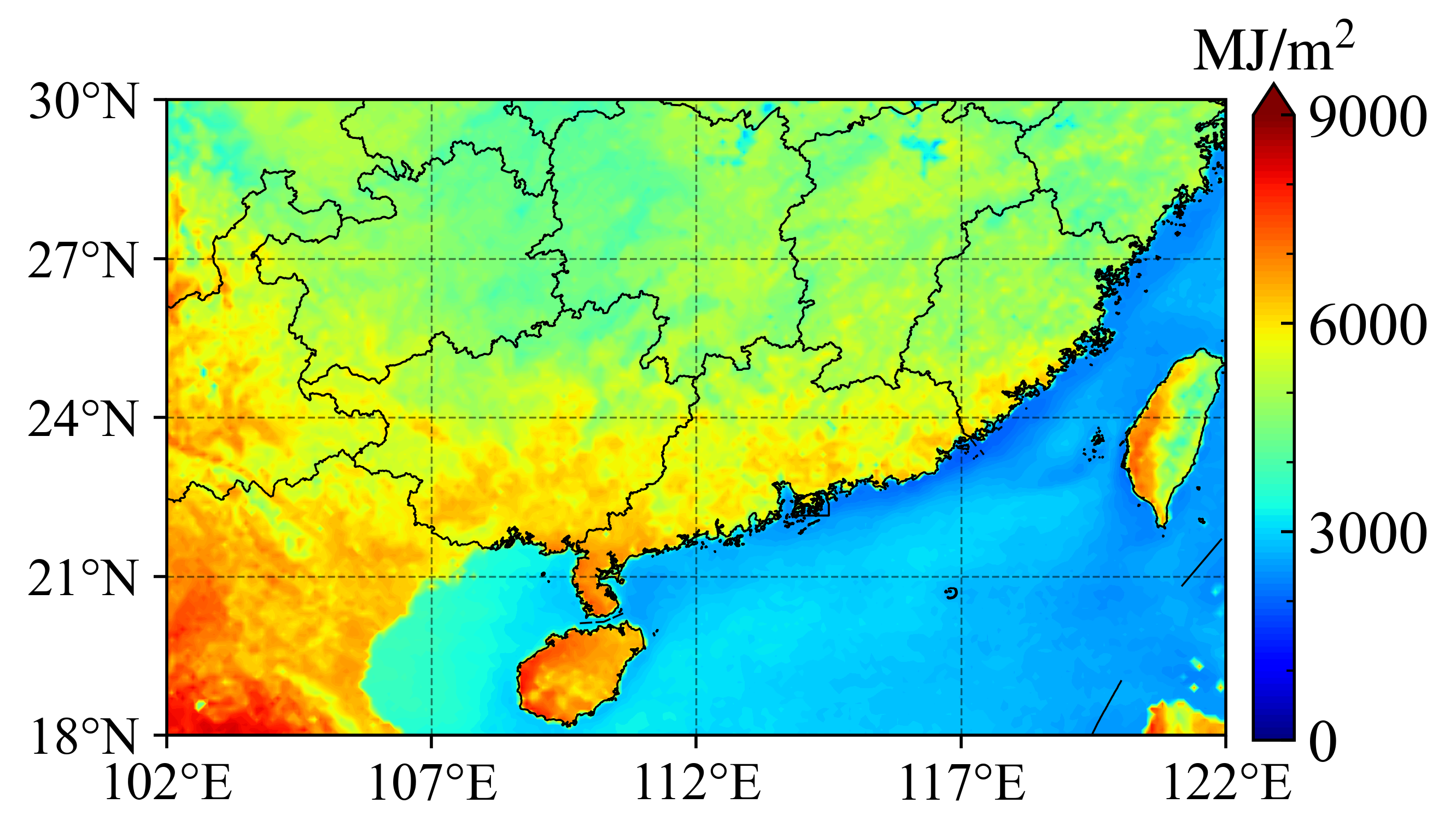}
    }
    \subcaptionbox{\label{fig:irrad_map_year:M1012135}}{
        \includegraphics[width=0.31\textwidth,trim=0 0 0 0,clip]{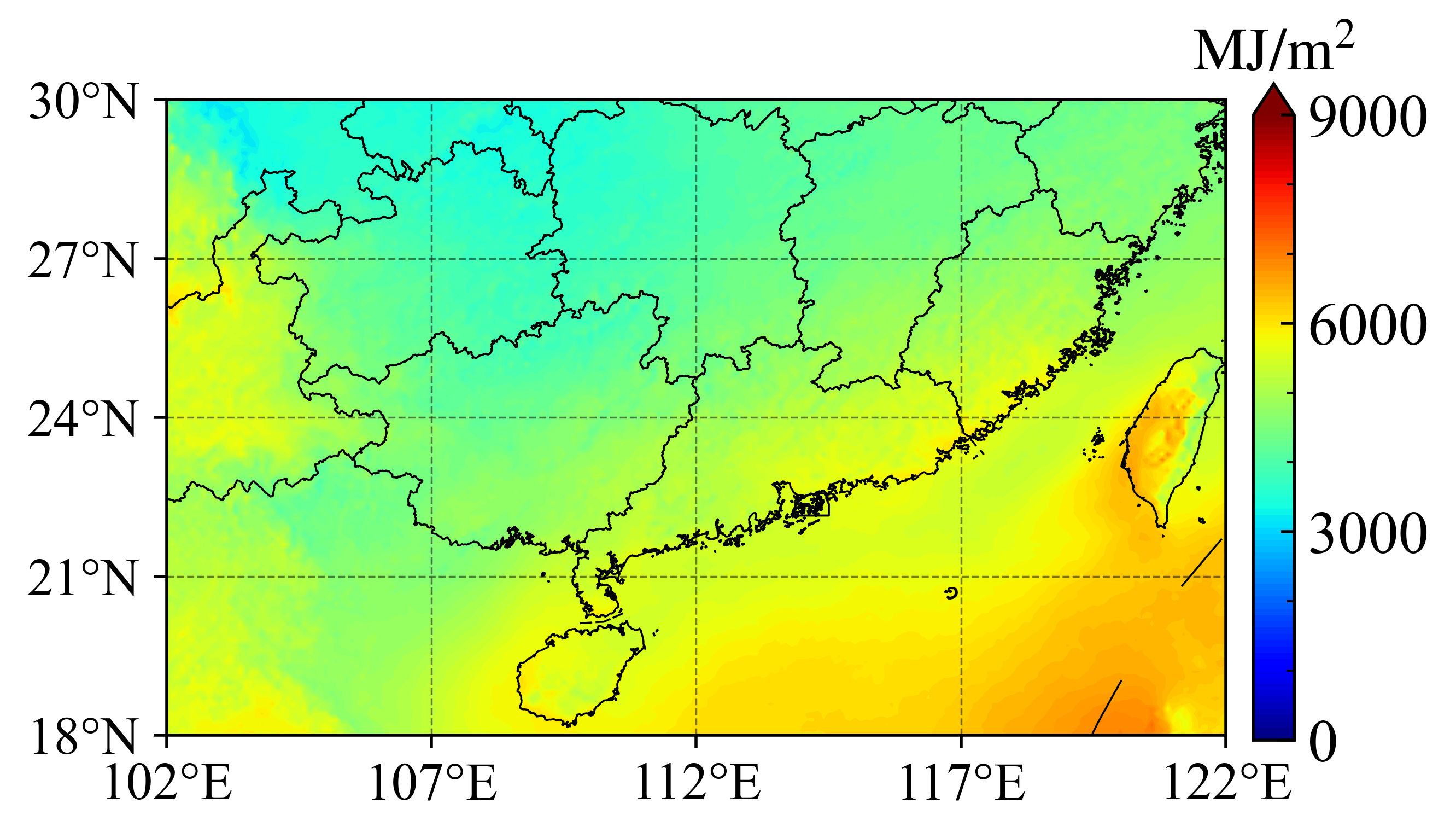}
    }
    \subcaptionbox{\label{fig:irrad_map_year:M1012136}}{
        \includegraphics[width=0.31\textwidth,trim=0 0 0 0,clip]{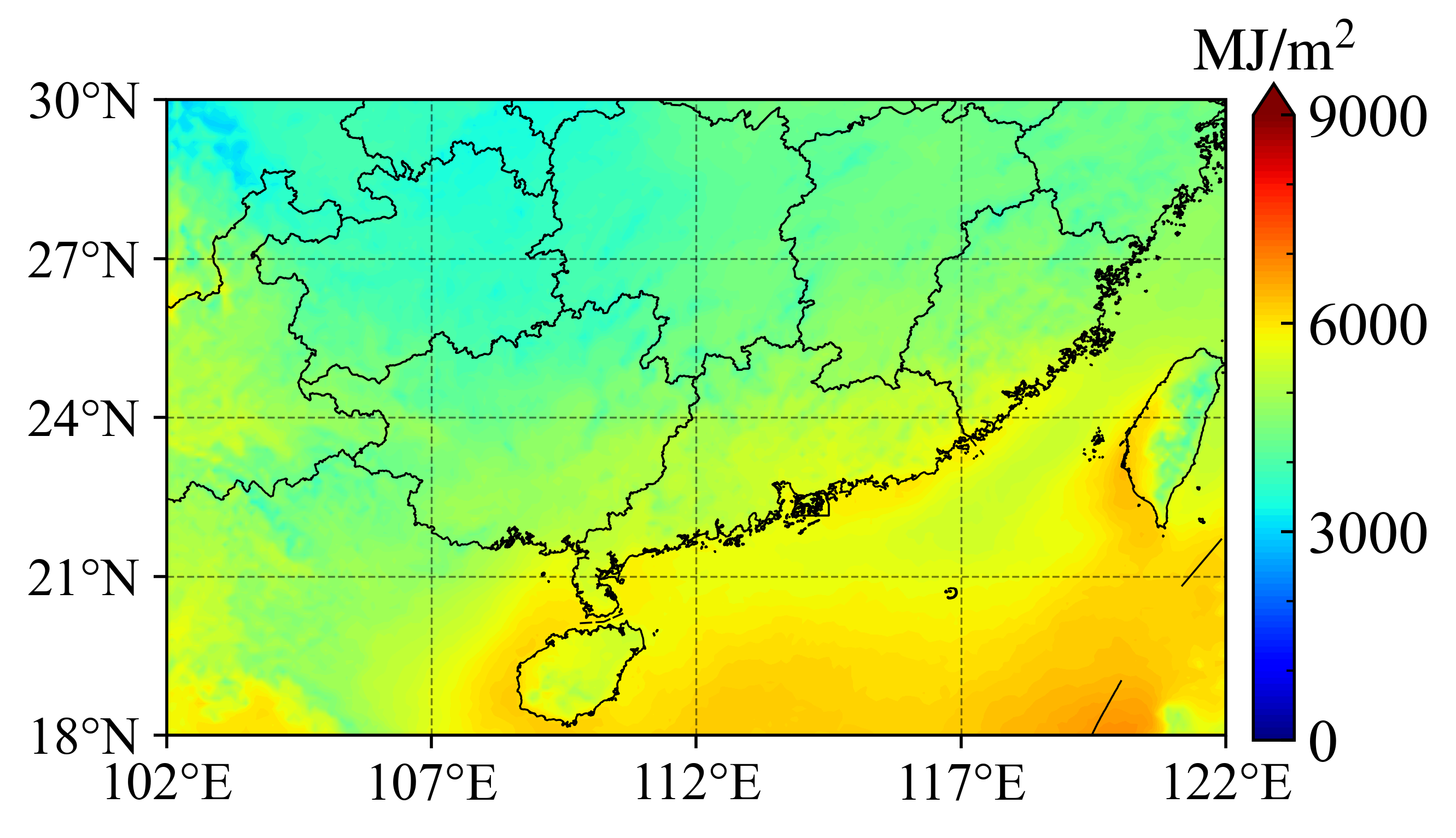}
    }
    \vspace{-1mm}
    \subcaptionbox{\label{fig:irrad_map_year:M1012137}}{
        \includegraphics[width=0.31\textwidth,trim=0 0 0 0,clip]{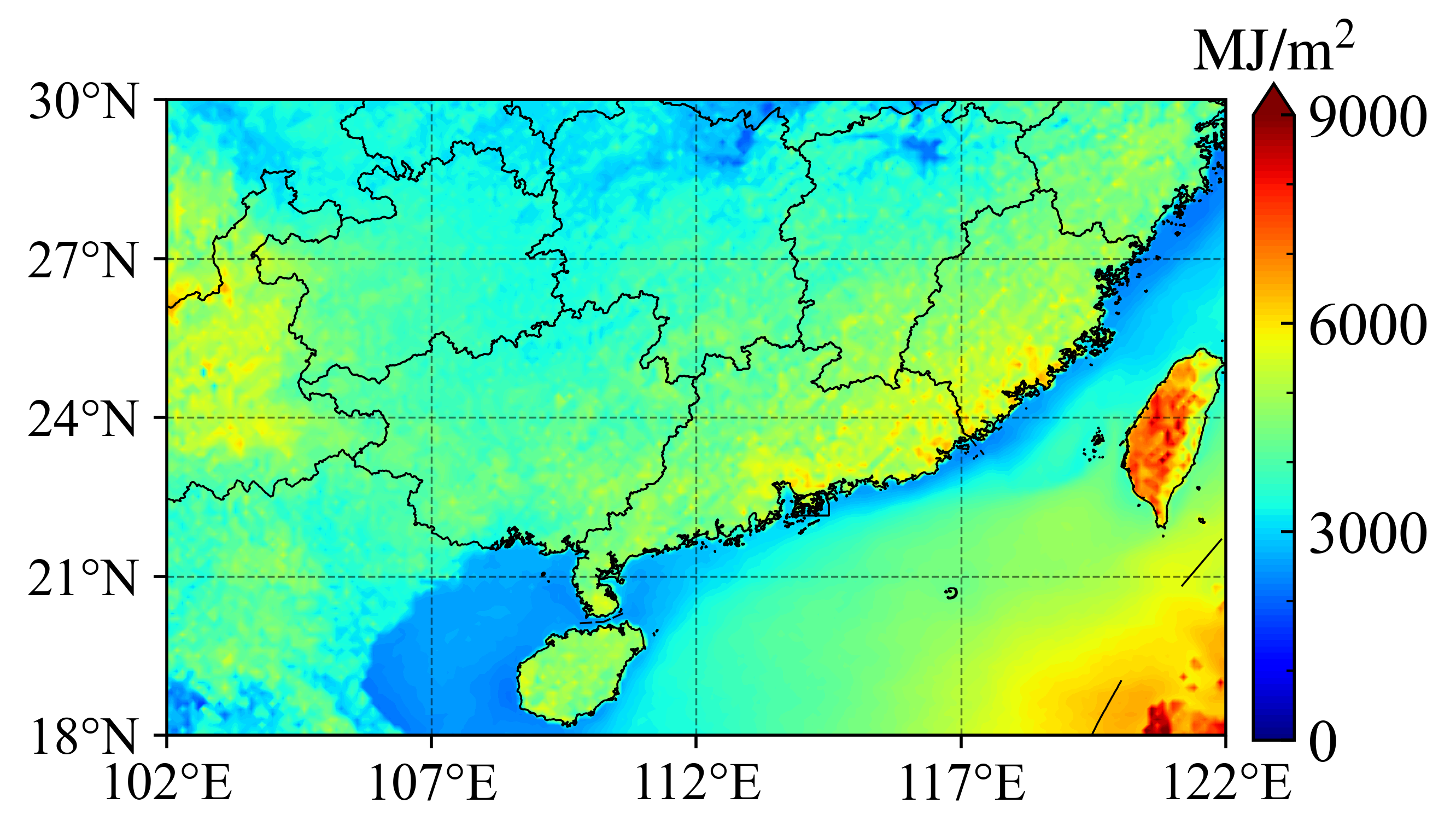}
    }
    \subcaptionbox{\label{fig:irrad_map_year:M1012138}}{
        \includegraphics[width=0.31\textwidth,trim=0 0 0 0,clip]{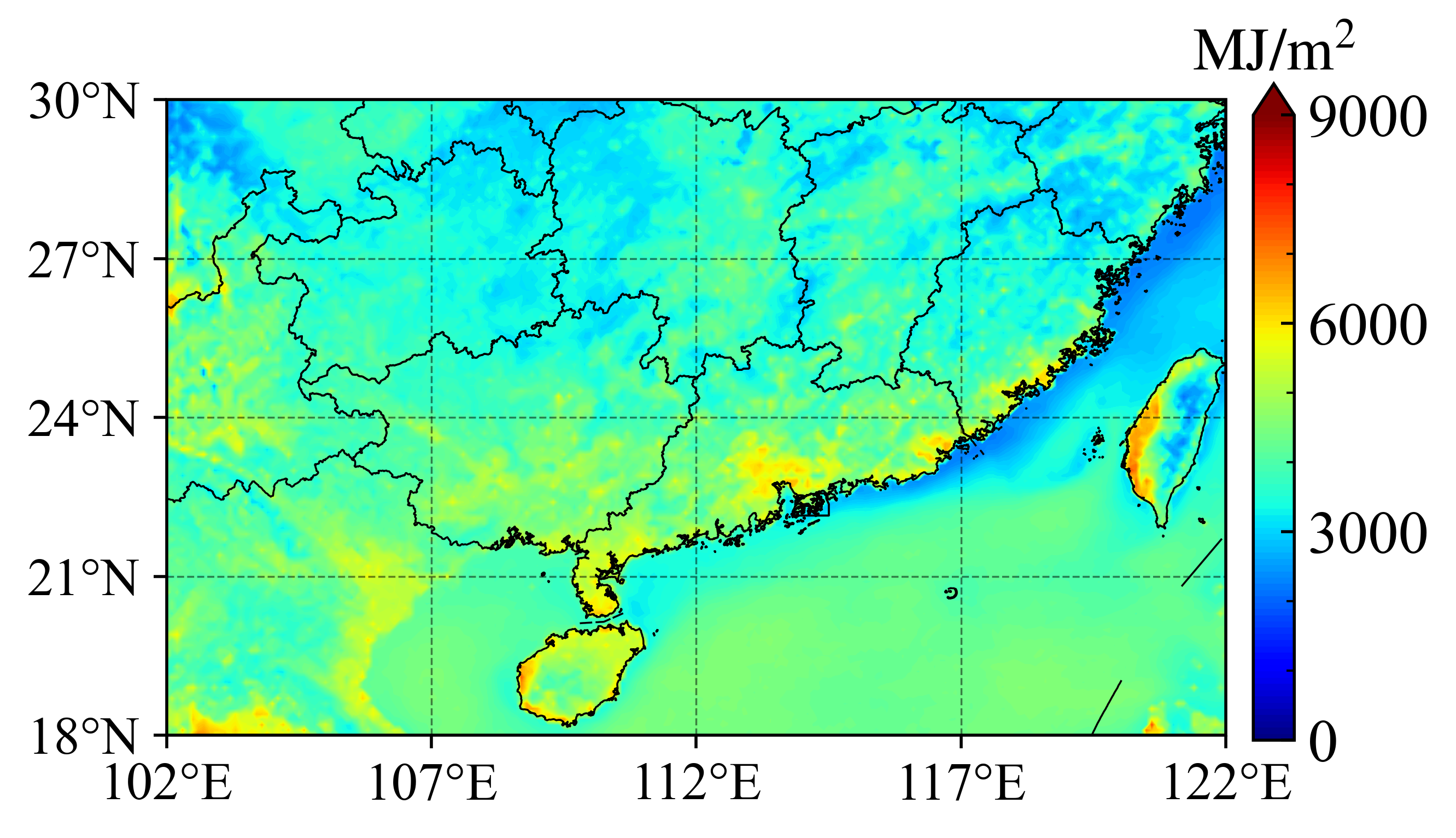}
    }
    \subcaptionbox{\label{fig:irrad_map_year:ERA5}}{
        \includegraphics[width=0.31\textwidth,trim=0 0 0 0,clip]{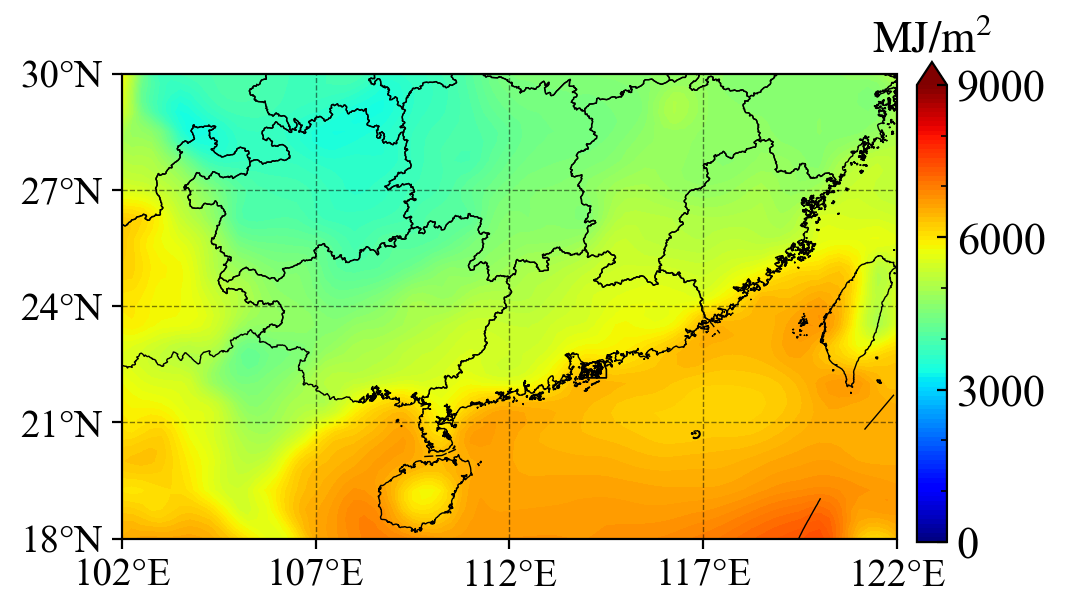}
    }
    \vspace{-1mm}
    \subcaptionbox{\label{fig:irrad_map_year:M1022035}}{
        \includegraphics[width=0.31\textwidth,trim=0 0 0 0,clip]{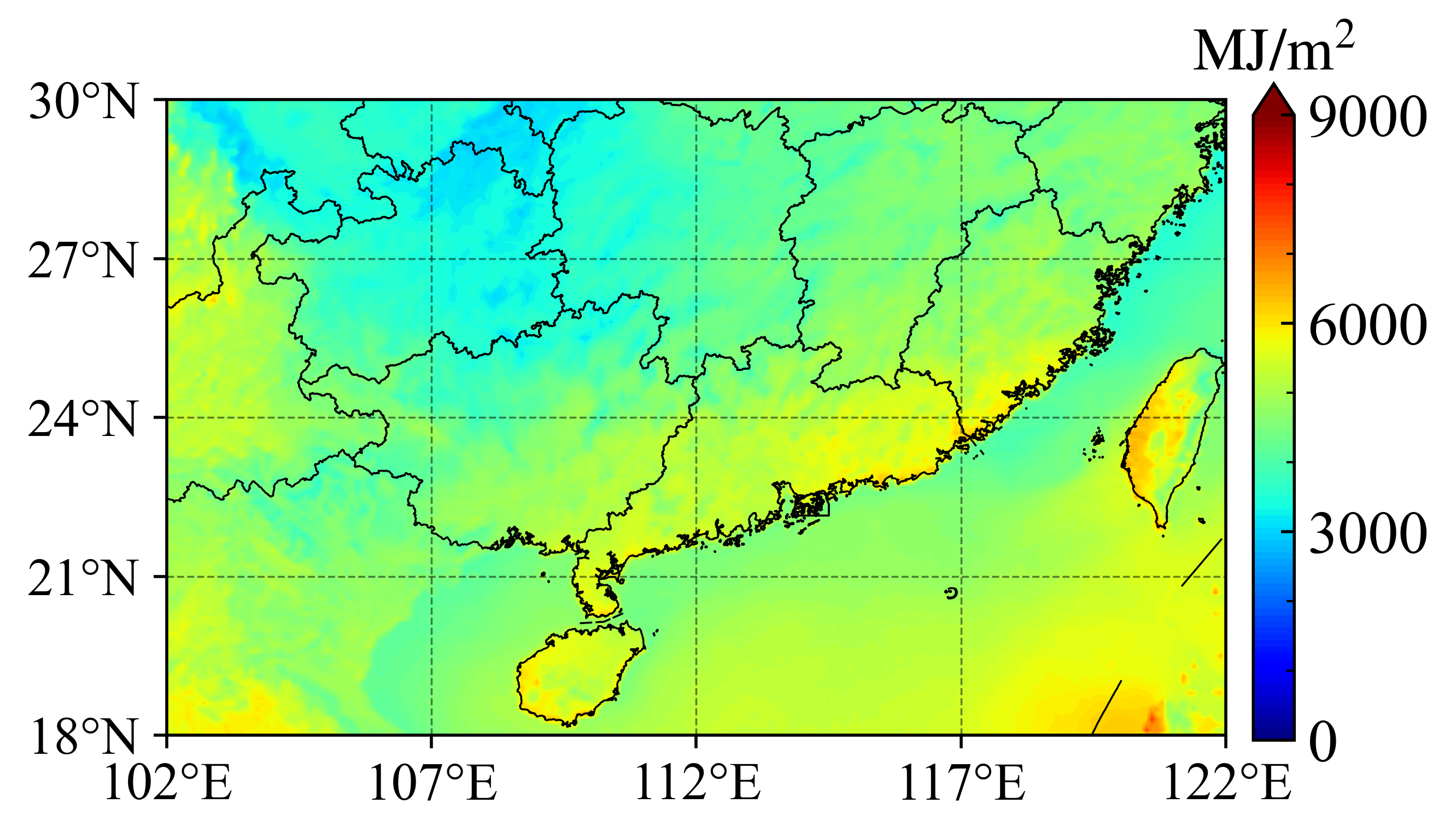}
    }
    \subcaptionbox{\label{fig:irrad_map_year:M1022036}}{
        \includegraphics[width=0.31\textwidth,trim=0 0 0 0,clip]{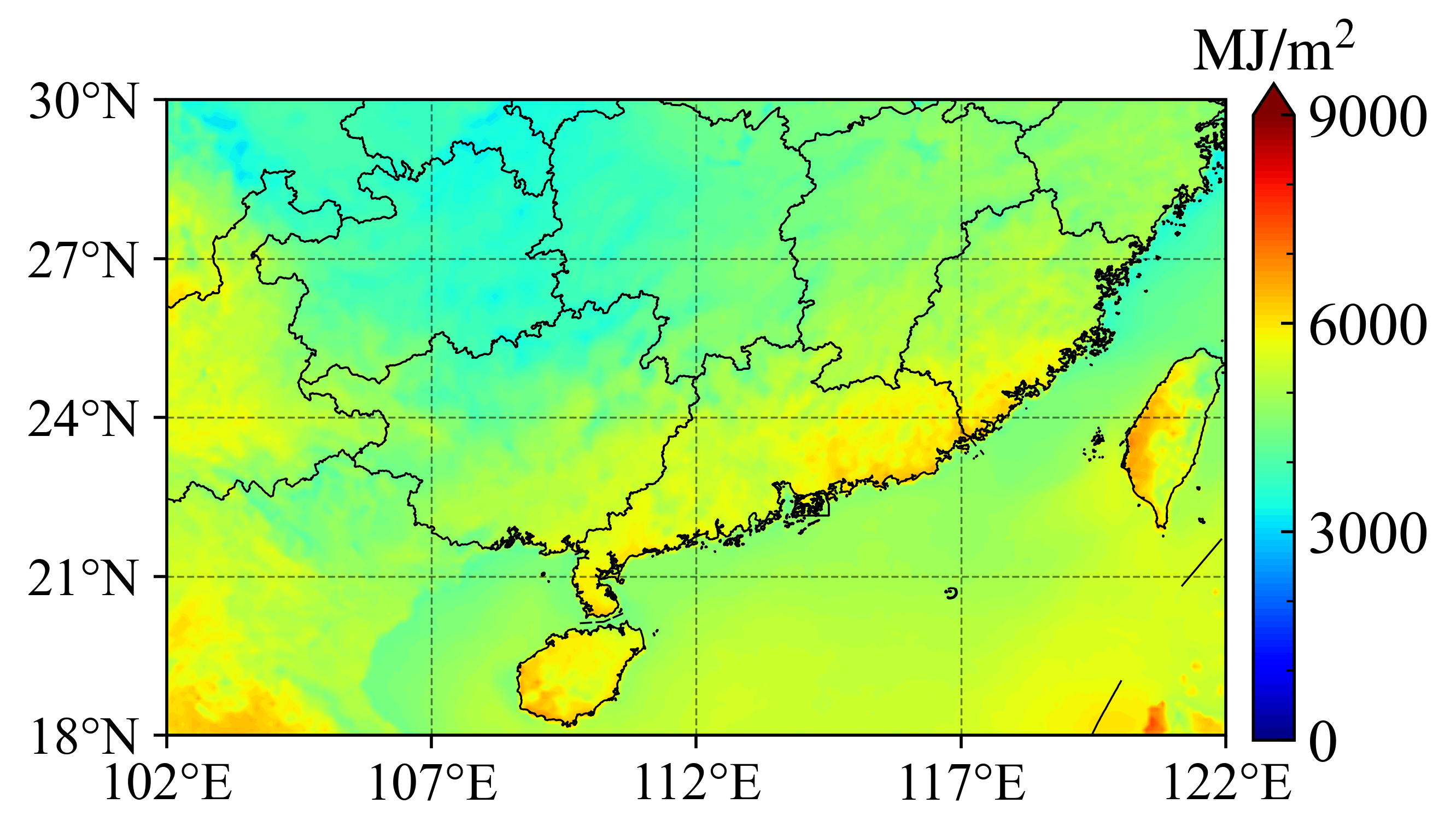}
    }
    \subcaptionbox{\label{fig:irrad_map_year:M1022037}}{
        \includegraphics[width=0.31\textwidth,trim=0 0 0 0,clip]{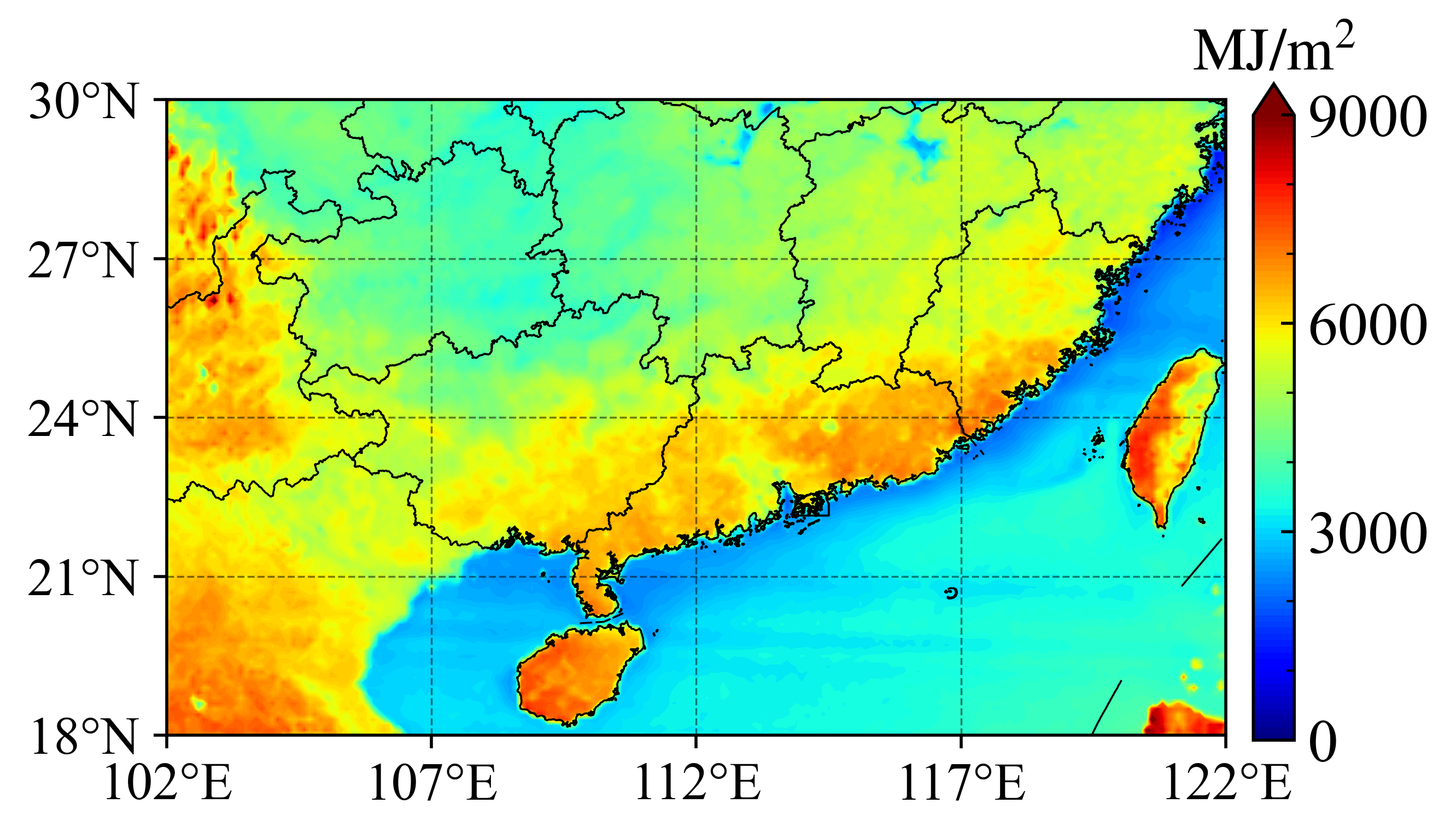}
    }
    \vspace{-1mm}
    \subcaptionbox{\label{fig:irrad_map_year:M1022038}}{
        \includegraphics[width=0.31\textwidth,trim=0 0 0 0,clip]{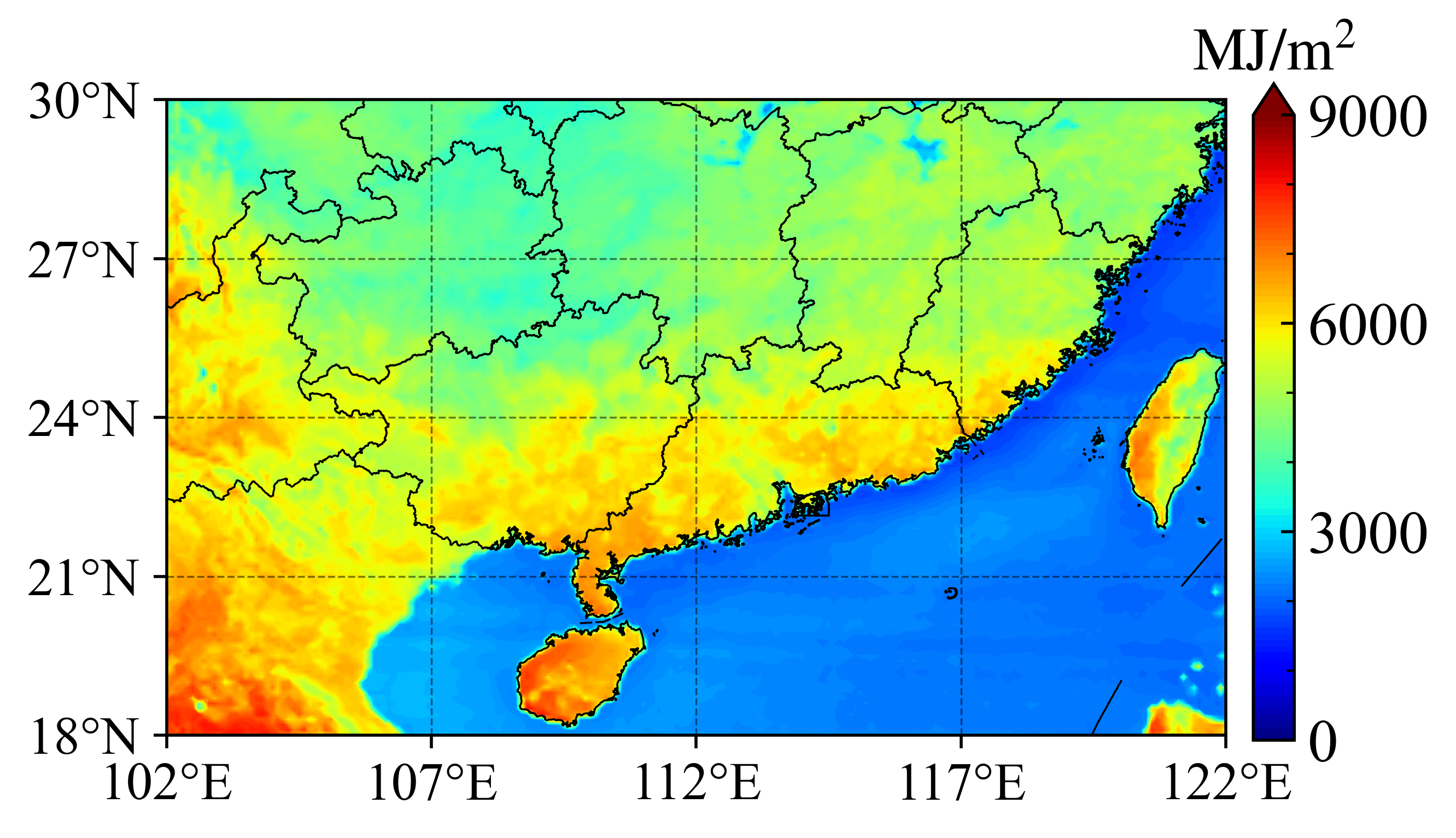}
    }
    \subcaptionbox{\label{fig:irrad_map_year:M1022135}}{
        \includegraphics[width=0.31\textwidth,trim=0 0 0 0,clip]{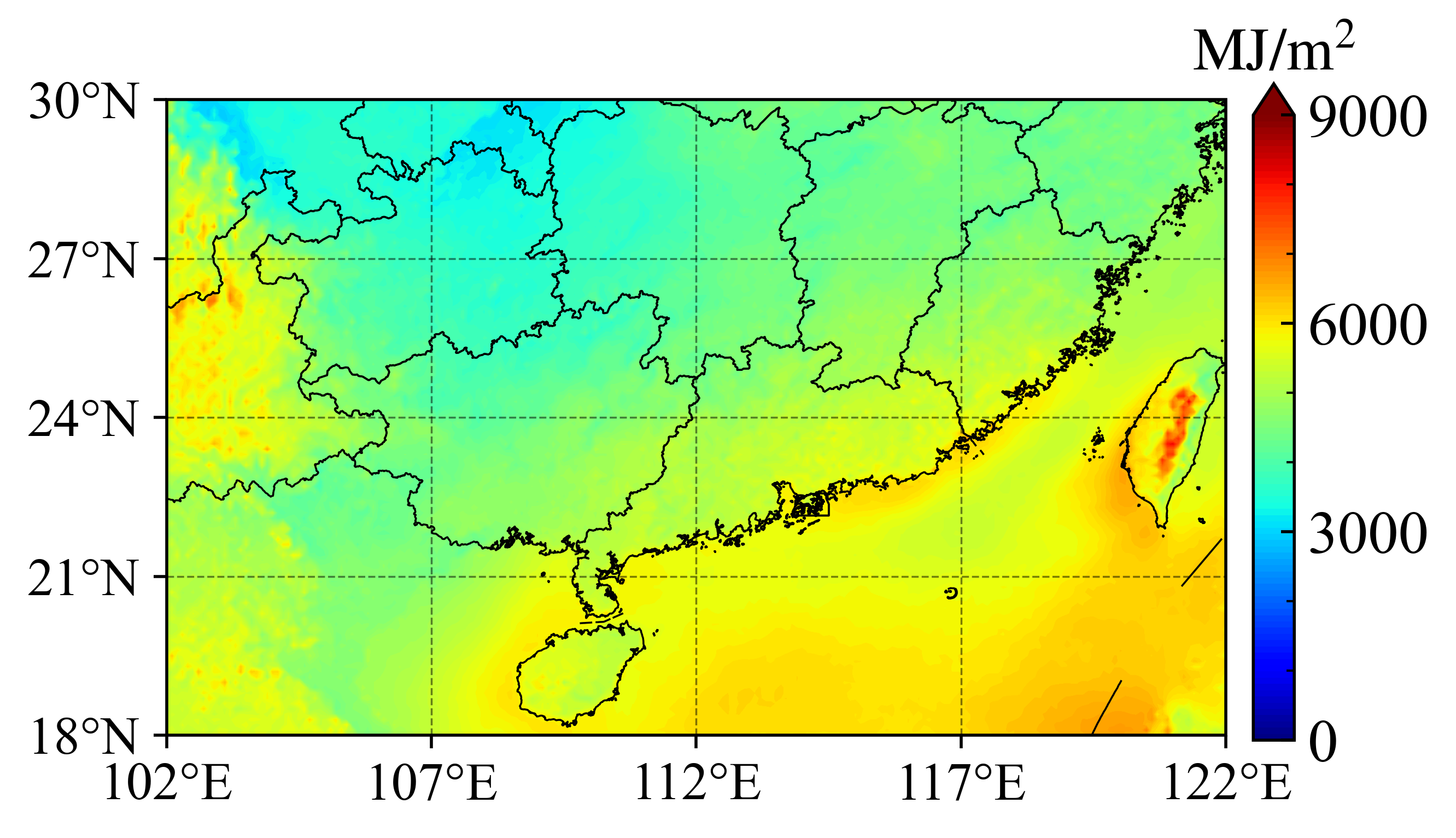}
    }
    \subcaptionbox{\label{fig:irrad_map_year:M1022136}}{
        \includegraphics[width=0.31\textwidth,trim=0 0 0 0,clip]{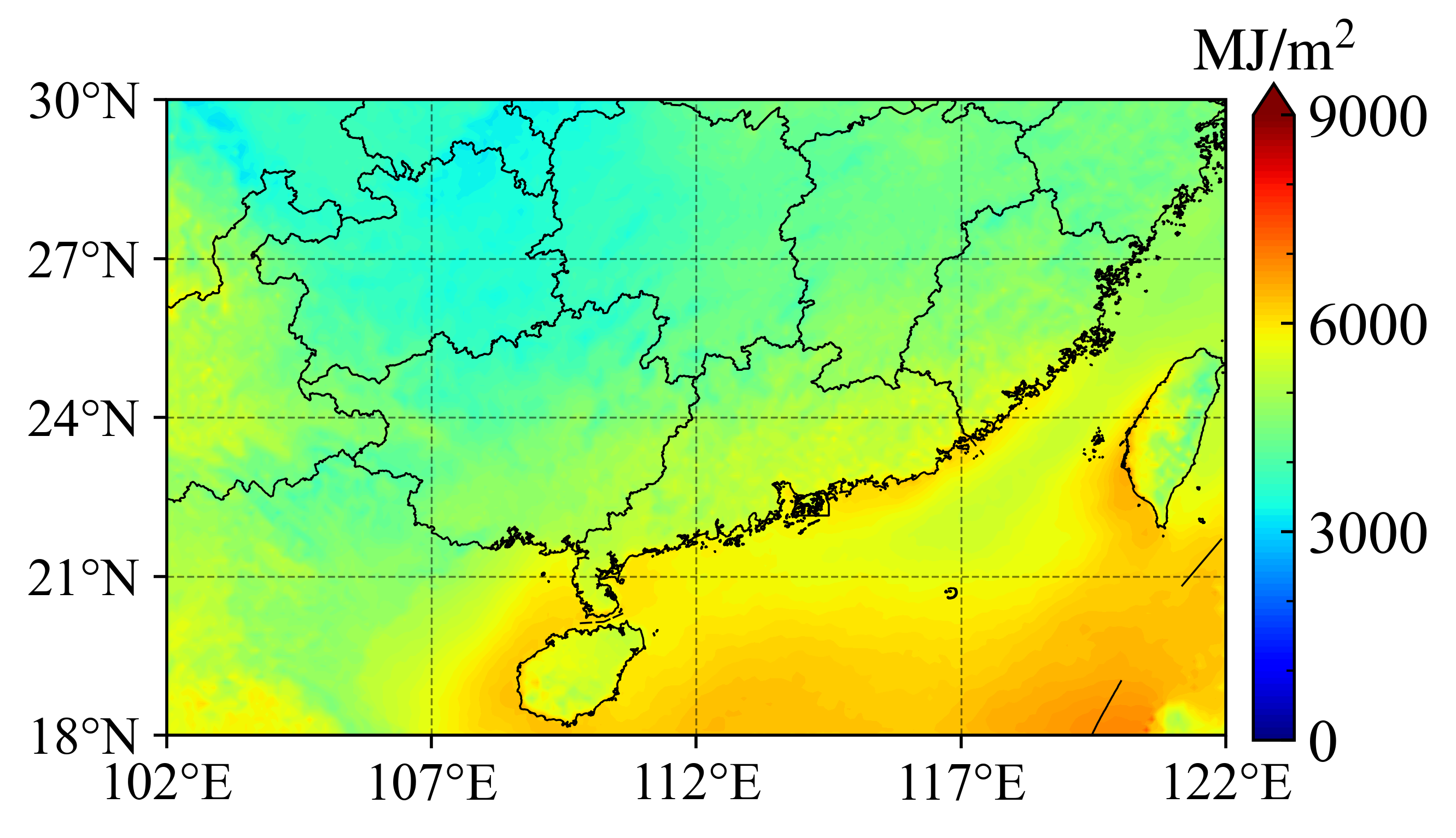}
    }
    \vspace{-1mm}
    \subcaptionbox{\label{fig:irrad_map_year:M1022137}}{
        \includegraphics[width=0.31\textwidth,trim=0 0 0 0,clip]{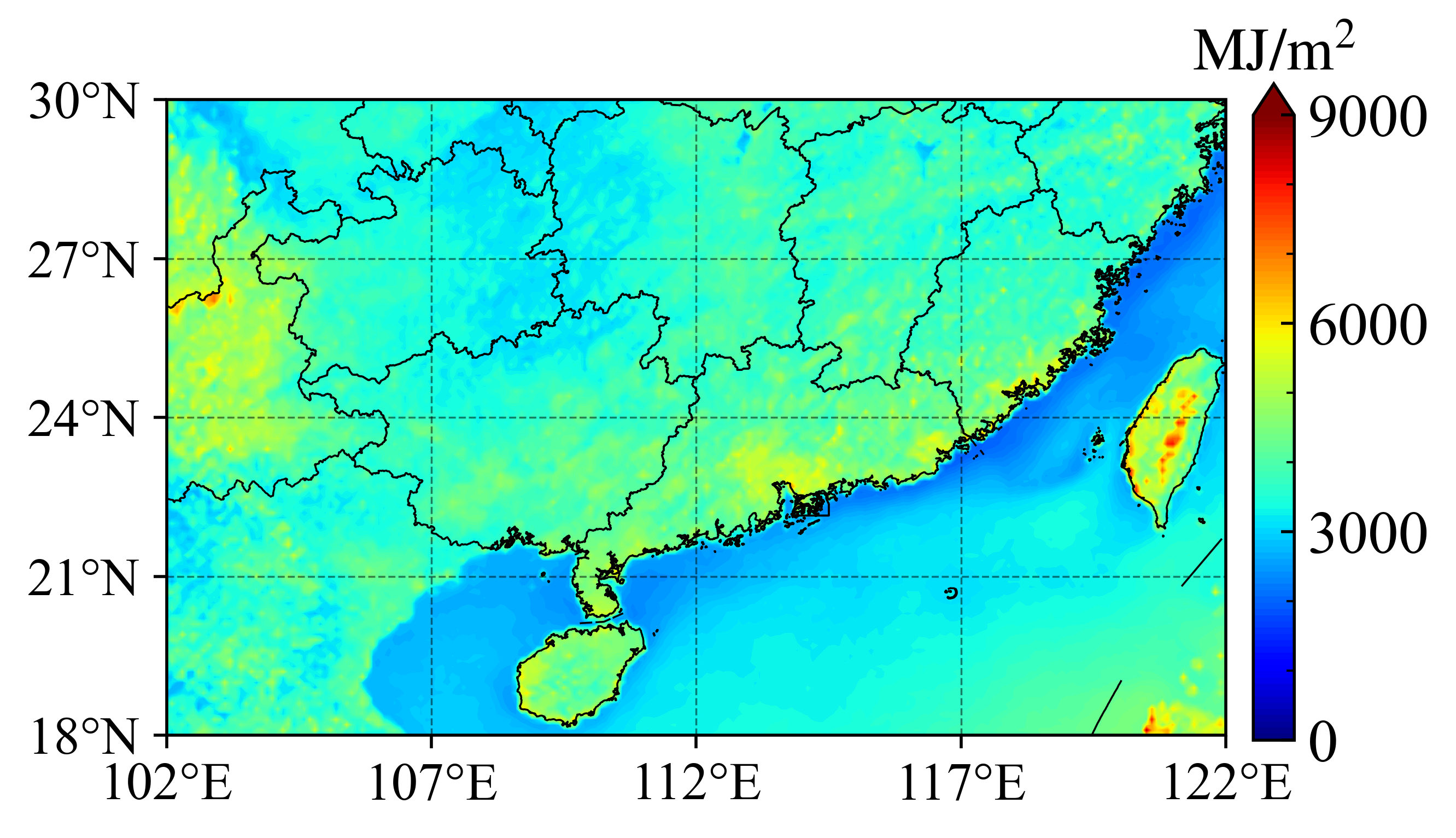}
    }
    \subcaptionbox{\label{fig:irrad_map_year:M1022138}}{
        \includegraphics[width=0.31\textwidth,trim=0 0 0 0,clip]{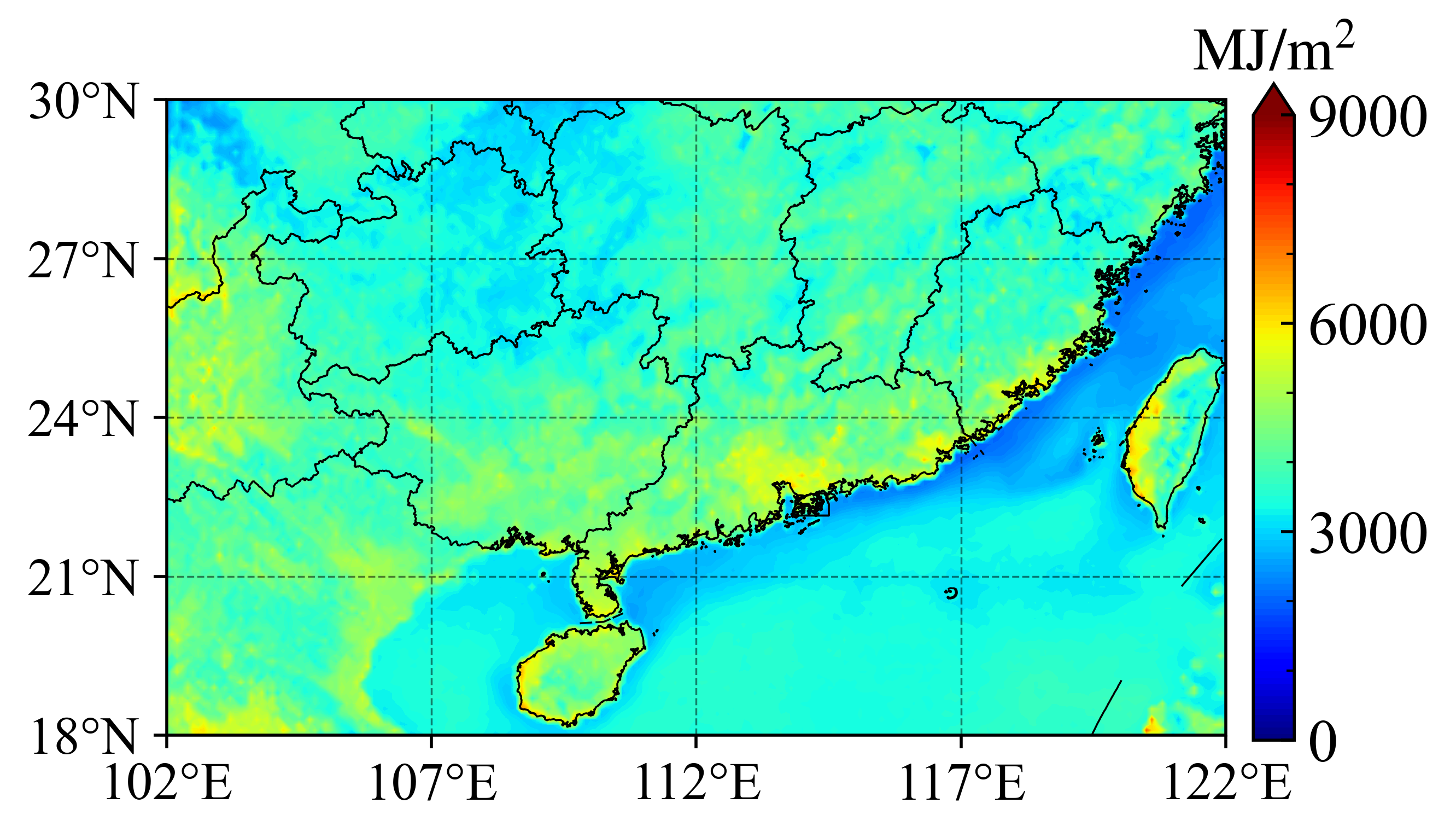}
    }
    \subcaptionbox{\label{fig:irrad_map_year:JAXA}}{
        \includegraphics[width=0.31\textwidth,trim=0 0 0 0,clip]{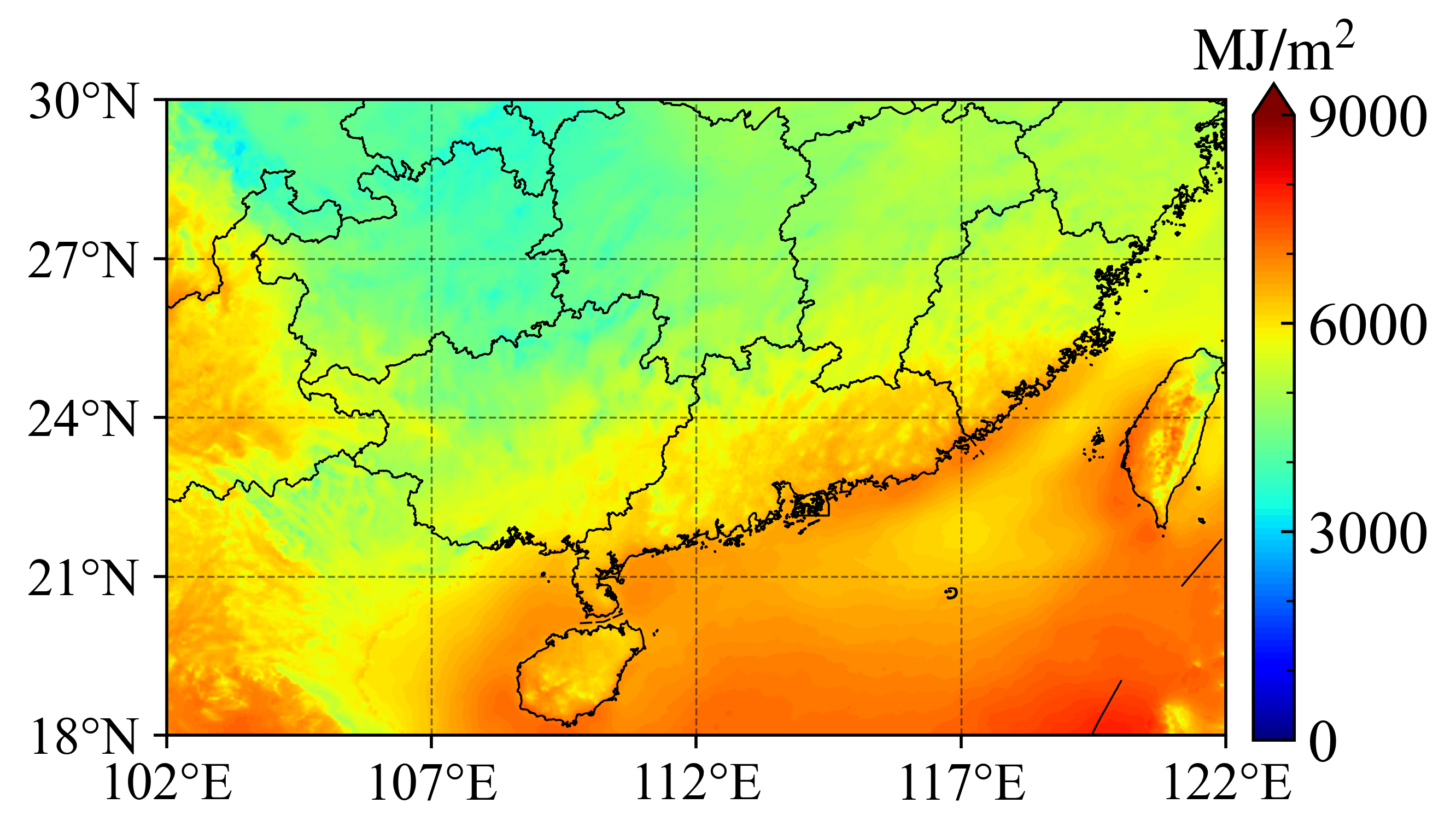}
    }
    \caption{Overall distributions of annual total solar energy estimated by (a) QIENet\_FC1, (b) QIENet\_FC2, (c) QIENet\_FC3, (d) QIENet\_FC4, (e) QIENet\_FC5, (f) QIENet\_FC6, (g) QIENet\_FC7, (h) QIENet\_FC8, (i) ERA5, (j) QIENet\_Conv1, (k) QIENet\_Conv2, (l) QIENet\_Conv3, (m) QIENet\_Conv4, (n) QIENet\_Conv5, (o) QIENet\_Conv6, (p) QIENet\_Conv7, (q) QIENet\_Conv8, and (r) the JAXA product from July 2020 to June 2021.}
    \label{fig:irrad_map_year}
\end{figure}

Hourly GHI estimates of the whole study area are reconstructed by QIENet, and the monthly, quarterly and annual solar energy are computed by integrating hourly GHI estimates.
Fig.~\ref{fig:irrad_map_year} depicts the overall distributions of annual total solar energy estimated by QIENet, the JAXA product and ERA5 from July 2020 to June 2021.
In Fig.~\ref{fig:irrad_map_year:M1012037}-\ref{fig:irrad_map_year:M1012038},~\ref{fig:irrad_map_year:M1012137}-\ref{fig:irrad_map_year:M1012138},~\ref{fig:irrad_map_year:M1022037}-\ref{fig:irrad_map_year:M1022038}, and~\ref{fig:irrad_map_year:M1022137}-\ref{fig:irrad_map_year:M1022138}, the annual solar energy is seriously underestimated in the ocean area and overestimated on land.
This phenomenon implies that the training process of QIENet cannot function without the time variable to update model parameters.
The fact that solar energy is underestimated also exists in the Beibu Gulf and the southeast coastal areas of China in Fig.~\ref{fig:irrad_map_year:M1012035}-\ref{fig:irrad_map_year:M1012036}, and~\ref{fig:irrad_map_year:M1022035}-\ref{fig:irrad_map_year:M1022036}.
The solar energy distributions of the JAXA product and ERA5 are overestimated, which is discussed in section~\ref{sub:Performance of QIENet}.
The solar energy distributions obtained by QIENet using BT of B07 and B11 - B15 as model inputs (QIENet\_FC6 and QIENet\_Conv6) are most reasonable, as shown in Fig.~\ref{fig:irrad_map_year:M1012136} and~\ref{fig:irrad_map_year:M1022136}.

\begin{figure}[!htbp]
    \vskip-0pt
    \centering
    \subcaptionbox{\label{fig:mapview1:M1022136}}{
        \begin{overpic}[abs,unit=1mm,width=0.96\textwidth,trim=0 0 0 0,clip]{
            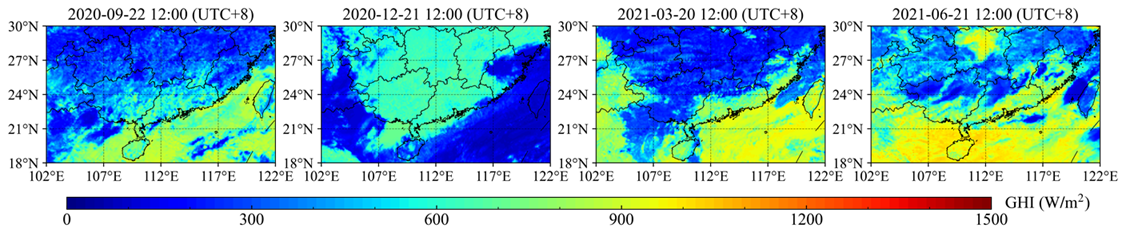
        }
        \put(0,0){
            \color{red}
            \linethickness{0.3mm}
            \polygon(25,11)(25,15)(33,15)(33,11)
            \polygon(149,19)(151,23)(154,23)(152,19)
        }
        \end{overpic}
    }
    \subcaptionbox{\label{fig:mapview1:JAXA}}{
        \begin{overpic}[abs,unit=1mm,width=0.96\textwidth,trim=0 0 0 0,clip]{
            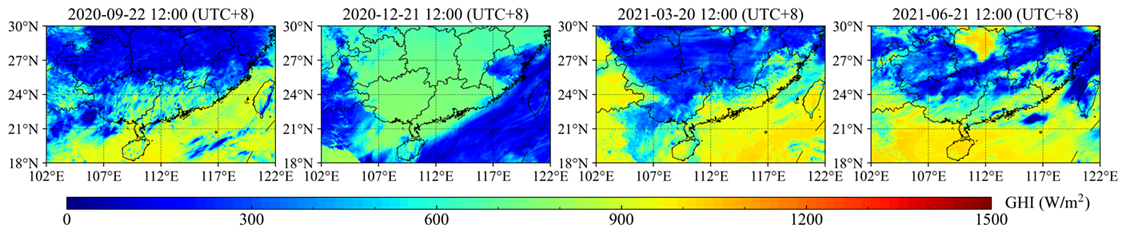
        }
        \put(0,0){
            \color{red}
            \linethickness{0.3mm}
            \polygon(25,11)(25,15)(33,15)(33,11)
            \polygon(149,19)(151,23)(154,23)(152,19)
        }
        \end{overpic}
    }
    \subcaptionbox{\label{fig:mapview1:ERA5}}{
        \begin{overpic}[abs,unit=1mm,width=0.96\textwidth,trim=0 0 0 0,clip]{
            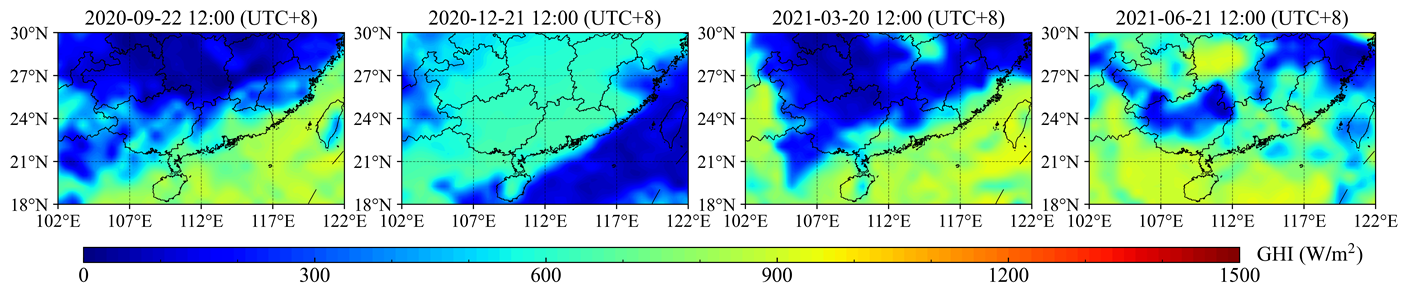
        }
        \put(0,0){
            \color{red}
            \linethickness{0.3mm}
            \polygon(25,11)(25,15)(33,15)(33,11)
            \polygon(149,19)(151,23)(154,23)(152,19)
        }
        \end{overpic}
    }
    \subcaptionbox{\label{fig:mapview1:AOT0400}}{
        \includegraphics[width=0.96\textwidth,trim=0 0 0 0,clip]{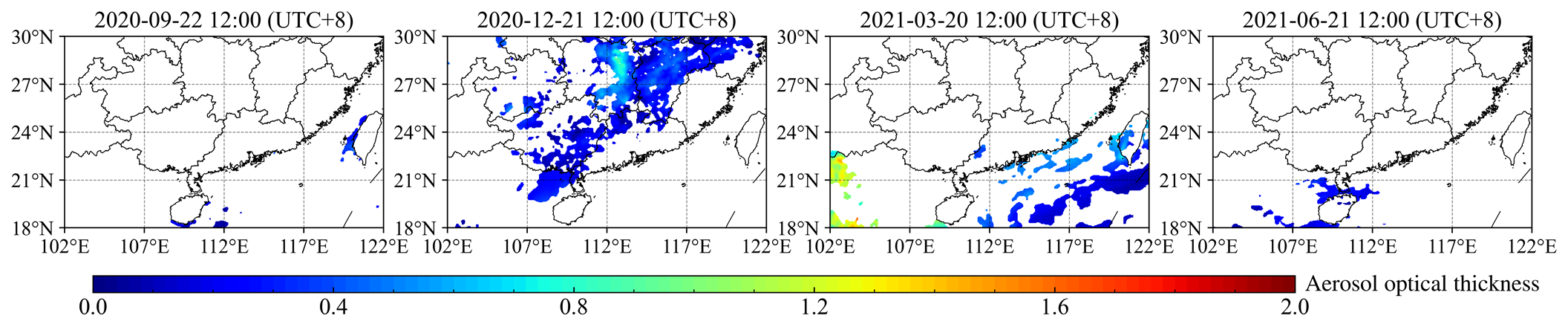}
    }
    \subcaptionbox{\label{fig:mapview1:CLOT0400}}{
        \begin{overpic}[abs,unit=1mm,width=0.96\textwidth,trim=0 0 0 0,clip]{
            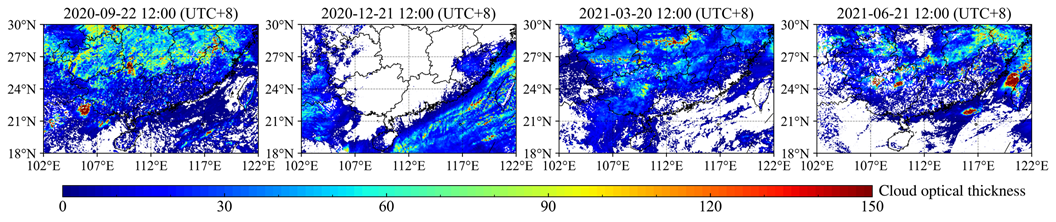
        }
        \put(0,0){
            \color{red}
            \linethickness{0.3mm}
            \polygon(25,11)(25,15)(33,15)(33,11)
            \polygon(149,19)(151,23)(154,23)(152,19)
        }
        \end{overpic}
    }
    \subcaptionbox{\label{fig:mapview1:cltype0400}}{
        \begin{overpic}[abs,unit=1mm,width=0.96\textwidth,trim=0 0 0 0,clip]{
            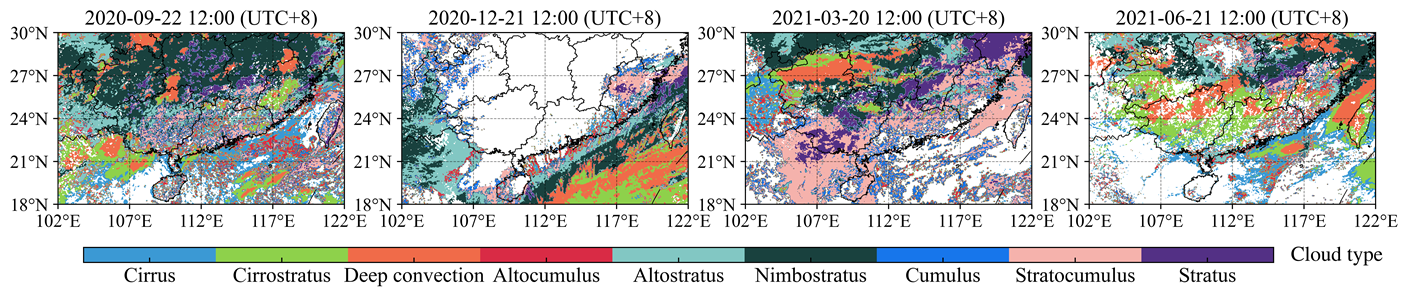
        }
        \put(0,0){
            \color{red}
            \linethickness{0.3mm}
            \polygon(25,11)(25,15)(33,15)(33,11)
            \polygon(149,19)(151,23)(154,23)(152,19)
        }
        \end{overpic}
    }
    \caption{GHI spatial distributions using (a) QIENet\_Conv6, (b) the JAXA product, and (c) ERA5 at 12:00 for the four solar terms, including the autumnal equinox (2020-09-22), the winter solstice (2020-12-21), the spring equinox (2021-03-20), and the summer solstice (2021-06-21), and the spatial distributions of (d) aerosol optical thickness, (e) cloud optical thickness, and (f) cloud type at 12:00 for the four solar terms.}
    \label{fig:mapview1}
\end{figure}

To further evaluate the performance of QIENet, the GHI spatial distributions using QIENet\_Conv6,the JAXA product, and ERA5 at 12:00 for the four solar terms, including the autumnal equinox (2020-09-22), the winter solstice (2020-12-21), the spring equinox (2021-03-20), and the summer solstice (2021-06-21), and the spatial distributions of aerosol optical thickness, cloud optical thickness and cloud type at 12:00, are shown in Fig.~\ref{fig:mapview1}.
Here, the Himawari L2/L3 gridded data cloud property and aerosol property are obtained from \href{https://www.eorc.jaxa.jp/ptree/index.html}{the JAXA Himawari Monitor P-Tree System}.
Cloud property and aerosol property have an important function in the absorption and transmission of radiation.
Therefore, the spatial distributions of aerosol optical thickness, cloud optical thickness and cloud type are used to verify the correctness of GHI estimates using QIENet\_Conv6.
For the autumnal equinox, QIENet\_Conv6 and the JAXA product can capture the effects of deep convection clouds on the sunlight reaching the sea surface in the area of 114° - 119° E and 18° - 21° N, but ERA5 does not.
For the summer solstice, deep convective clouds appear in the Taiwan Strait.
At the same time, the optical thickness of clouds in the Taiwan Strait is large, which will diminish the ability of sunlight to penetrate clouds.
Results of QIENet\_Conv6 and the JAXA product well indicate the blocking effect of such clouds on sunlight, but the results of ERA5 do not.
Overall, from the analysis of the above results, both QIENet\_Conv6 and the JAXA product have a good ability to estimate GHI with high accuracy.

\begin{figure}[!htbp]
    \centering
    \includegraphics[width=0.96\textwidth,trim=0 0 0 0,clip]{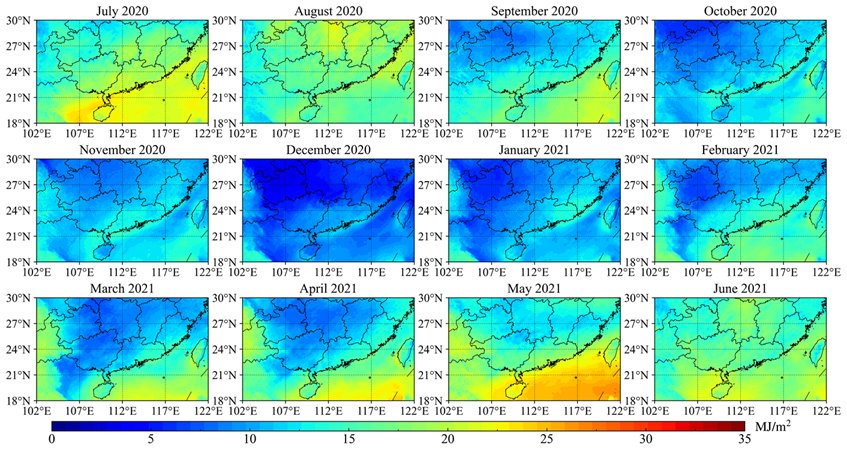}
    \caption{Overall distributions of monthly mean daily solar energy integrated from hourly GHI estimates using QIENet\_Conv6.}
    \label{fig:irrad_map_month}
\end{figure}

\begin{figure}[!htbp]
    \vskip-0pt
    \vspace{-5mm}
    \centering
    \includegraphics[width=0.96\textwidth,trim=0 0 0 0,clip]{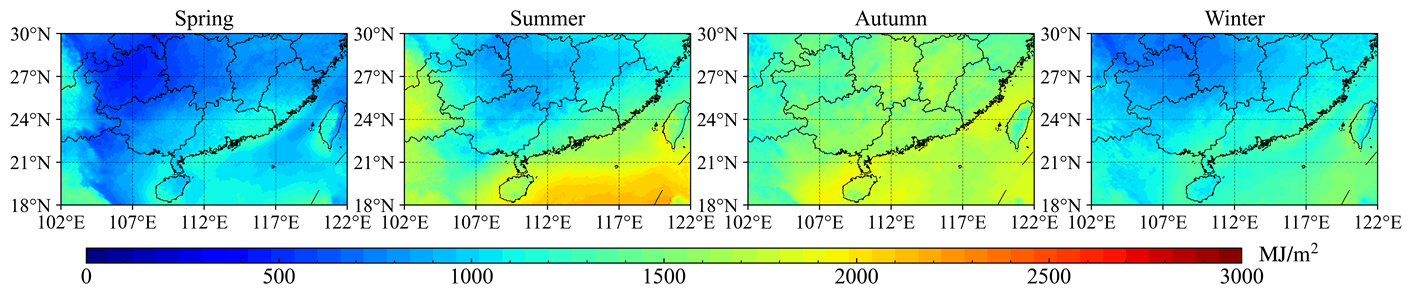}
    \caption{Spatial distributions of quarterly total solar energy estimated by QIENet\_Conv6 for spring (December, January, and February), summer (March, April, and May), autumn (June, July, and August), and winter (September, October, and November) from July 2020 to June 2021.}
    \label{fig:irrad_map_season}
\end{figure}

Monthly and quarterly solar energy distributions are now introduced in detail.
The overall distributions of monthly mean daily solar energy integrated from hourly GHI estimates of model QIENet\_Conv6 are displayed in Fig.~\ref{fig:irrad_map_month}.
During the annual cycle, solar energy levels are higher from May to August than those in other months, while the solar energy level in December is the lowest.
The overall solar energy distributions in terms of season are shown by using QIENet\_Conv6 in Fig.~\ref{fig:irrad_map_season}.
The estimated solar energy levels in autumn and summer are higher than those in spring and winter, and the solar energy level is the lowest in spring and the highest in autumn.
Moreover, our research results suggest abundant solar energy in southeast China and Yunnan Province, which is used to generate clean electricity by installing photovoltaic panels.

\section{Conclusions}
\label{section:5}
In this study, a quantitative irradiance estimation network, named QIENet, is proposed and applied to estimate hourly GHI values in the study area (102° - 122° E, 18° - 30° N) by employing RNN, whose basic unit structure adopts the widely used LSTM.
Standard RNN and convolutional RNN are applied into QIENet to build QIENet-FCRNN and QIENet-ConvRNN, respectively.
By exploring the effects of the satellite Himawari-8 spectral channels, time and geographical information, on QIENet in section~\ref{sub:the_effects_of_input_variables}, there are three main findings:
(1) Satellite spectral channels B07 and B11-B15 are selected as model inputs instead of all satellite spectral channels, whose performance is almost the same, and it is evident that the role of B01-B06 can be replaced by information from other channels;
(2) The input variable of time is more important than geographical information, and the performance of QIENet becomes poor without the input variable of time;
(3) Under the same input conditions, QIENet-ConvRNN outperforms QIENet-FCRNN, which indicates that spatial fusion realized by convolution operation is vital for QIENet.

Hourly GHI estimates of the whole study area are reconstructed by QIENet, and the monthly, quarterly and annual solar energy with spatial resolution $0.02^{\circ}\times0.02^{\circ}$ (about $\mathrm{2km \times 2km}$) are calculated by integrating hourly GHI estimates from July 2020 to June 2021.
From the perspective of the distribution of annual solar energy, using remote sensing data from the satellite spectral channels B07 and B11-B15 and time as model inputs to train QIENet in this study is recommended.
QIENet\_Conv6 can achieve similar performance to the JAXA product, but does not overestimate ground observations.
Meanwhile, QIENet\_Conv6 can also reduce RMSE by 27.51\%/18.00\%, increase $\mathrm{R^{2}}$ by 20.17\%/9.42\%, and increase r by 8.69\%/3.54\% compared with ERA5/NSRDB.
More importantly, compared with ERA5, our model QIENet\_Conv6 is able to capture the impact of various clouds on hourly GHI estimates.
For long-term research, the impact of satellite spectral channel combinations on GHI estimation, GHI estimation algorithms with higher temporal resolution, regional GHI forecast, and GHI estimation embedding physical models to improve accuracy will be explored in the future.

\section*{CRediT authorship contribution statement}
Longfeng Nie: formal analysis, methodology, visualization, writing - original draft. Yuntian Chen: funding acquisition, writing - review and editing. Dongxiao Zhang: supervision, funding acquisition, writing - review and editing. Xinyue Liu: writing - review and editing. Wentian Yuan: data acquisition, formal analysis.

\section*{Declaration of Competing Interest}
The authors declare that they have no known competing financial interests or personal relationships that could have appeared to influence the work reported in this paper.

\section*{Data availability}
The specific network structure settings and the trained network parameters are available at the GitHub page \href{https://github.com/rsai0/PMD/tree/main/QIENetV1_0_0}{https://github.com/rsai0/PMD/tree/main/QIENetV1\_0\_0}. Further updates will follow.

\section*{Acknowledgements}
This work was supported and partially funded by the Shenzhen Key Laboratory of Natural Gas Hydrates (Grant No. ZDSYS20200421111201738), the National Center for Applied Mathematics Shenzhen, the SUSTech-Qingdao New Energy Technology Research Institute, the Major Key Project of PCL (Grant No. PCL2022A05), and the National Natural Science Foundation of China (Grant No. 62106116).

\appendix
\section{}

\setcounter{table}{0}
\renewcommand{\thetable}{A\arabic{table}}   

\setcounter{figure}{0}
\renewcommand{\thefigure}{A\arabic{figure}}

\begin{table}[h]
    \caption{The spectral channels of the satellite Himawari-8.}
    \centering
    \resizebox{0.96\linewidth}{!}{
        \begin{tabular}{cccccc}
            \hline
            Satellite channel & Central wavelength (µm) & Bandwidth (µm) & Variable    & Resolution (km) & Primary application                        \\ \hline
            B01               & 0.455                   & 0.05           & Albedo (\%) & 1.0             & Aerosol                                    \\
            B02               & 0.510                   & 0.02           & Albedo (\%) & 1.0             & Aerosol                                    \\
            B03               & 0.645                   & 0.03           & Albedo (\%) & 0.5             & Fog and low cloud                          \\
            B04               & 0.860                   & 0.02           & Albedo (\%) & 1.0             & Aerosol and vegetation                     \\
            B05               & 1.610                   & 0.02           & Albedo (\%) & 2.0             & Cloud phase                                \\
            B06               & 2.260                   & 0.02           & Albedo (\%) & 2.0             & Particle size                              \\
            B07               & 3.850                   & 0.22           & BT (K)      & 2.0             & Fog, low cloud, and forest fire            \\
            B08               & 6.250                   & 0.37           & BT (K)      & 2.0             & Upper level moisture                       \\
            B09               & 6.950                   & 0.12           & BT (K)      & 2.0             & Mid-upper level moisture                   \\
            B10               & 7.350                   & 0.17           & BT (K)      & 2.0             & Mid-level moisture                         \\
            B11               & 8.600                   & 0.32           & BT (K)      & 2.0             & SO2 and cloud phase                        \\
            B12               & 9.630                   & 0.18           & BT (K)      & 2.0             & Ozone content                              \\
            B13               & 10.45                   & 0.30           & BT (K)      & 2.0             & Information of cloud top and cloud imagery \\
            B14               & 11.20                   & 0.20           & BT (K)      & 2.0             & Sea surface temperature and cloud imagery  \\
            B15               & 12.35                   & 0.30           & BT (K)      & 2.0             & Sea surface temperature and cloud imagery  \\
            B16               & 13.30                   & 0.20           & BT (K)      & 2.0             & Cloud top height                           \\ \hline
        \end{tabular}
    }
    \label{table:AHI_info}
\end{table}

\begin{figure}[H]
    \centering
    \subcaptionbox{\label{fig:irrad_Albedo1}}{
        \includegraphics[width=.46\textwidth]{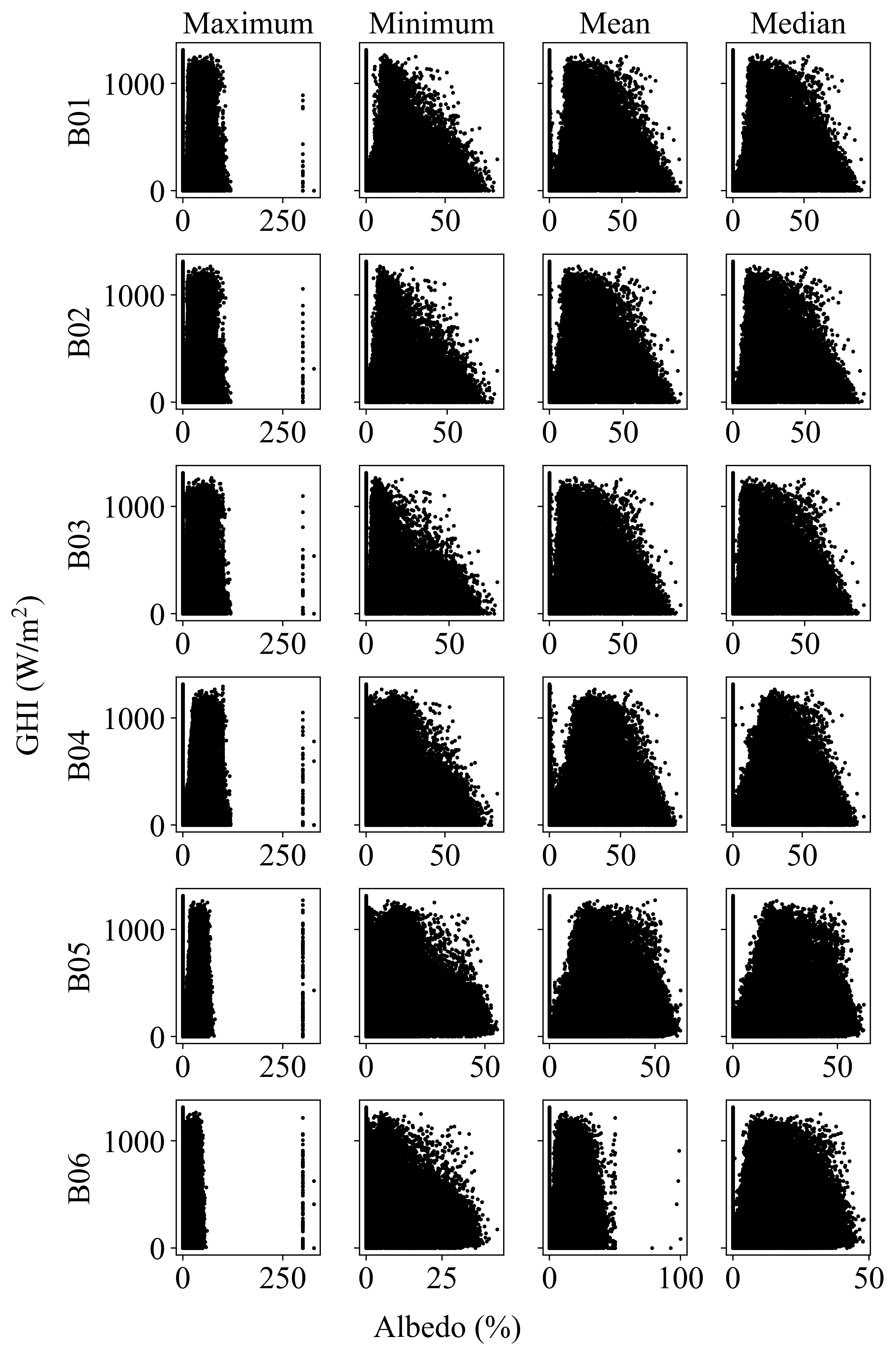}
    }
    \subcaptionbox{\label{fig:irrad_Albedo2}}{
        \includegraphics[width=.46\textwidth]{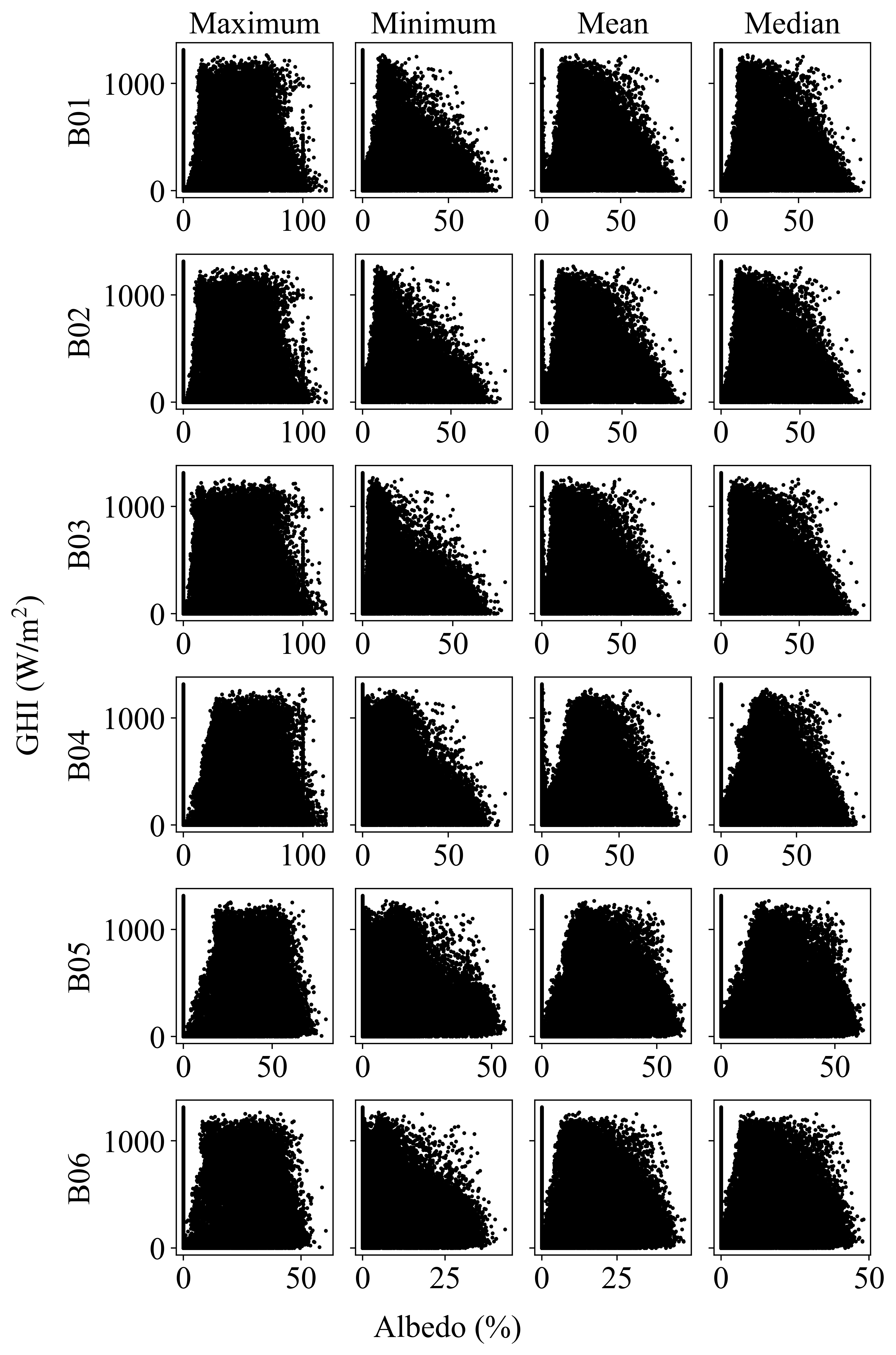}
    }
    \caption{The relationship distributions among hourly GHI observations and the maximum, minimum, mean and median values for each channel of the satellite image slices. (a) albedo of the original samples, (b) albedo of the corrected samples, (c) BT of the original samples, and (d) BT of the corrected samples.}
    \label{fig:irrad_Albedo_TBB}
\end{figure}
\begin{figure}[H]
    \centering
    \ContinuedFloat
    \subcaptionbox{\label{fig:irrad_TBB1}}{
        \includegraphics[width=.46\textwidth]{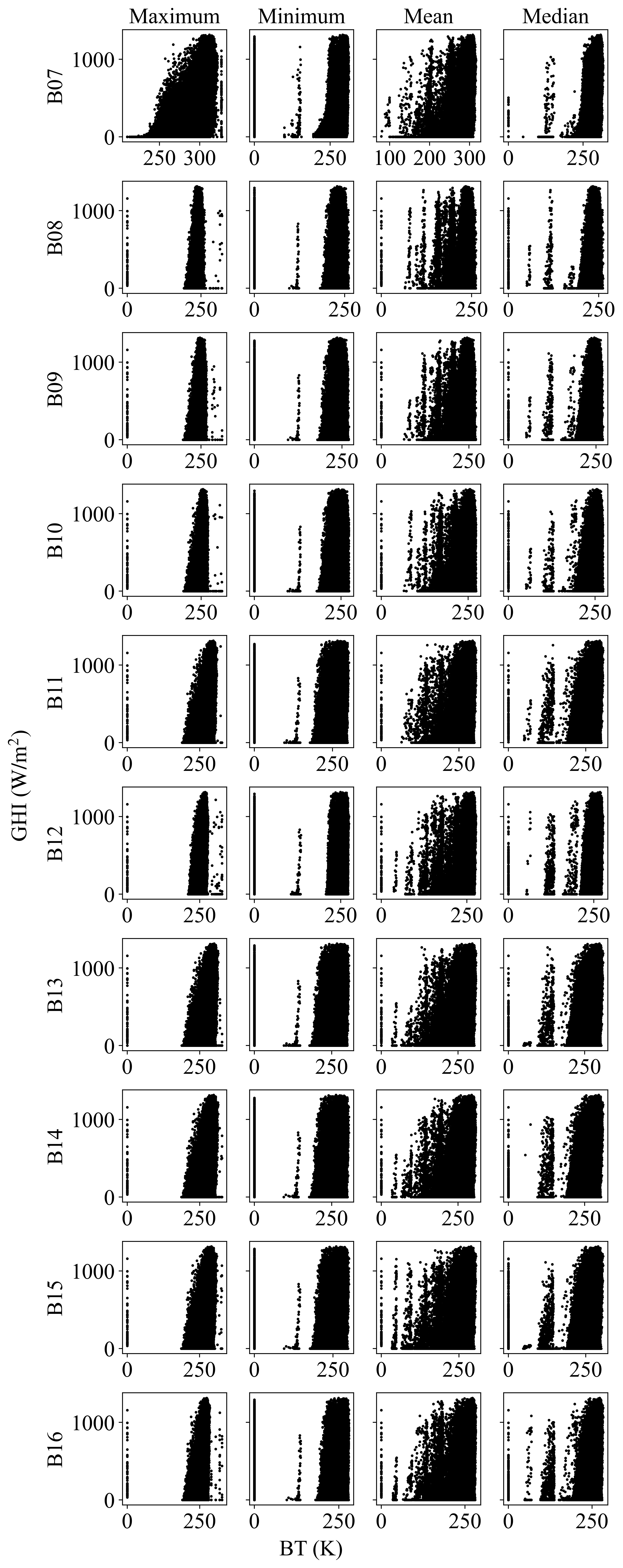}
    }
    \subcaptionbox{\label{fig:irrad_TBB2}}{
        \includegraphics[width=.46\textwidth]{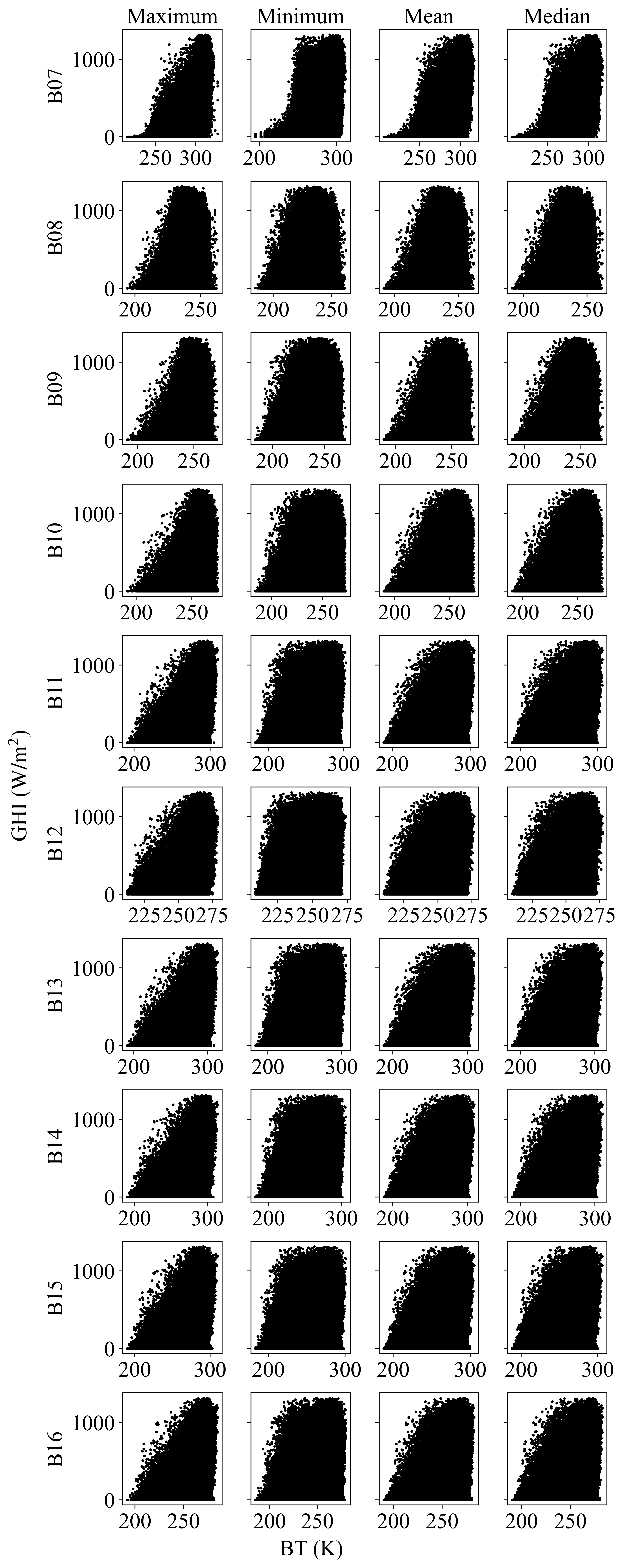}
    }
    \caption{Continued}
\end{figure}
\setcounter{subfigure}{0}

\begin{figure}[H]
    \centering
    \includegraphics[width=.75\textwidth]{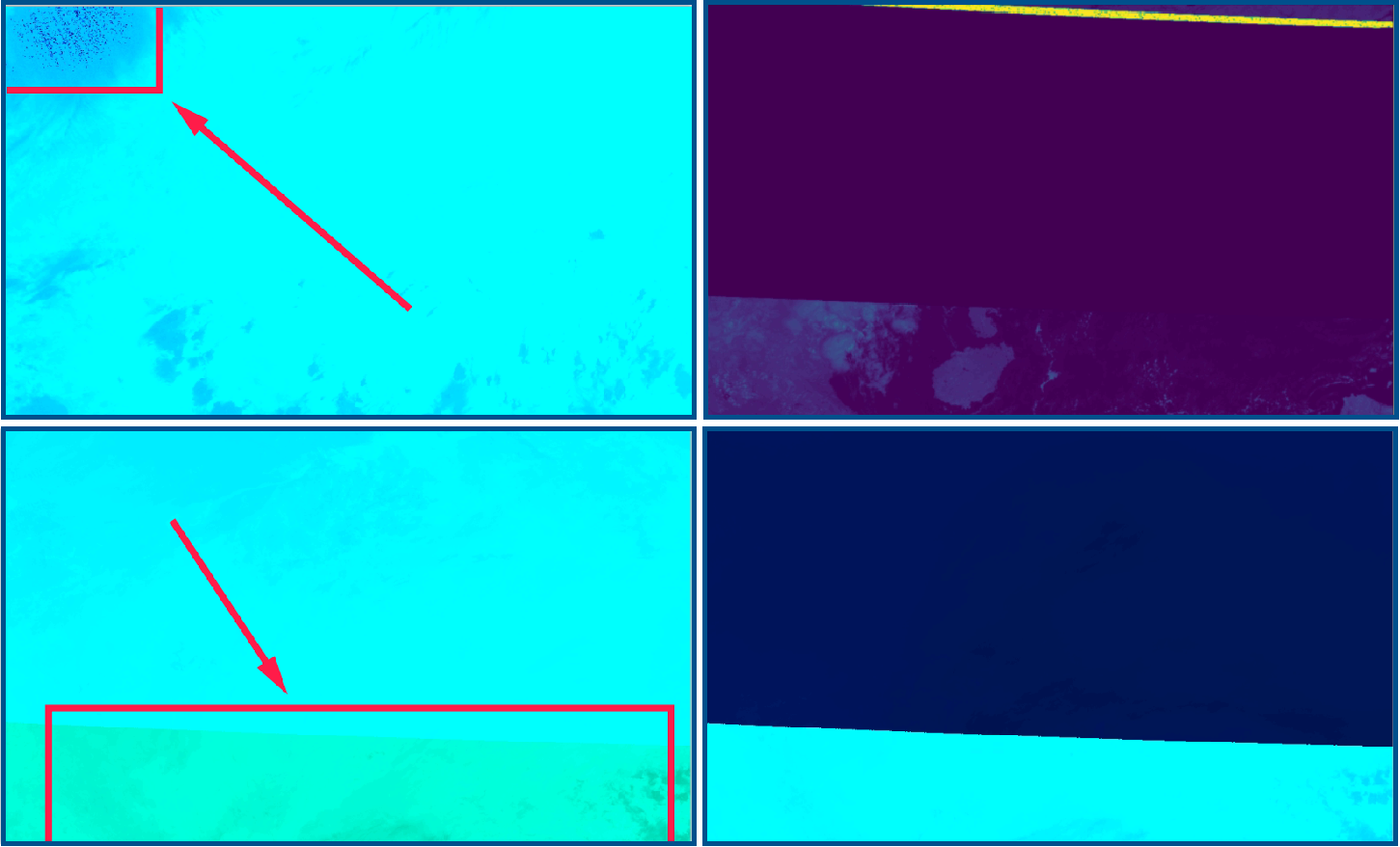}
    \caption{Examples of problematic satellite image}
    \label{fig:satellite_problem}
\end{figure}

\begin{table}[H]
    \caption{Different models of QIENet.}
    \centering
    \resizebox{0.7\linewidth}{!}{
        \begin{tabular}{ccccccc}
            \hline
            \multirow{2}{*}{Model} &  & \multicolumn{3}{c}{Input variables ($\blacksquare$: used, $\square$: not used)} &                & \multirow{2}{*}{Network type}                     \\ \cline{3-5}
                                &  & Satellite spectral channels                 & Time           & Geographic information      &  &                \\ \hline
            QIENet\_FC1          &  & B01 - B16                     & $\blacksquare$ & $\blacksquare$              &  & QIENet-FCRNN   \\
            QIENet\_FC2          &  & B01 - B16                     & $\blacksquare$ & $\square$                   &  & QIENet-FCRNN   \\
            QIENet\_FC3          &  & B01 - B16                     & $\square$      & $\blacksquare$              &  & QIENet-FCRNN   \\
            QIENet\_FC4          &  & B01 - B16                     & $\square$      & $\square$                   &  & QIENet-FCRNN   \\
            QIENet\_FC5          &  & B07, B11 - B15             & $\blacksquare$ & $\blacksquare$              &  & QIENet-FCRNN   \\
            QIENet\_FC6          &  & B07, B11 - B15             & $\blacksquare$ & $\square$                   &  & QIENet-FCRNN   \\
            QIENet\_FC7          &  & B07, B11 - B15             & $\square$      & $\blacksquare$              &  & QIENet-FCRNN   \\
            QIENet\_FC8          &  & B07, B11 - B15             & $\square$      & $\square$                   &  & QIENet-FCRNN   \\
            QIENet\_Conv1          &  & B01 - B16                     & $\blacksquare$ & $\blacksquare$              &  & QIENet-ConvRNN \\
            QIENet\_Conv2          &  & B01 - B16                     & $\blacksquare$ & $\square$                   &  & QIENet-ConvRNN \\
            QIENet\_Conv3          &  & B01 - B16                     & $\square$      & $\blacksquare$              &  & QIENet-ConvRNN \\
            QIENet\_Conv4          &  & B01 - B16                     & $\square$      & $\square$                   &  & QIENet-ConvRNN \\
            QIENet\_Conv5          &  & B07, B11 - B15             & $\blacksquare$ & $\blacksquare$              &  & QIENet-ConvRNN \\
            QIENet\_Conv6          &  & B07, B11 - B15             & $\blacksquare$ & $\square$                   &  & QIENet-ConvRNN \\
            QIENet\_Conv7          &  & B07, B11 - B15             & $\square$      & $\blacksquare$              &  & QIENet-ConvRNN \\
            QIENet\_Conv8          &  & B07, B11 - B15             & $\square$      & $\square$                   &  & QIENet-ConvRNN \\ \hline
        \end{tabular}
    }
    \label{table:model_inputs}
\end{table}

\begin{table}[H]
    \caption{Evaluation indicators of QIENet\_FC1, QIENet\_Conv1, QIENet\_Conv6, ERA5, JAXA, and NSRDB on the test dataset.}
    \centering
    \resizebox{0.5\linewidth}{!}{
        \begin{tabular}{ccccccc}
            \hline
            Model                          & Index                     & k=1    & k=2    & k=3    & k=4    & k=5    \\ \hline
            \multirow{4}{*}{QIENet\_FC1}   & RMSE ($\mathrm{W/m^{2}}$) & 98.16  & 98.18  & 98.18  & 98.23  & 98.15  \\
                                           & MBE ($\mathrm{W/m^{2}}$)  & -1.32  & -0.98  & 0.33   & -1.21  & -1.15  \\
                                           & $\mathrm{R^{2}}$          & 0.835  & 0.834  & 0.834  & 0.834  & 0.835  \\
                                           & r                         & 0.914  & 0.914  & 0.914  & 0.913  & 0.914  \\ \cline{2-7}
            \multirow{4}{*}{QIENet\_Conv1} & RMSE ($\mathrm{W/m^{2}}$) & 93.20  & 93.59  & 93.34  & 93.41  & 93.19  \\
                                           & MBE ($\mathrm{W/m^{2}}$)  & -1.86  & -0.87  & -1.99  & -2.86  & -1.87  \\
                                           & $\mathrm{R^{2}}$          & 0.851  & 0.850  & 0.850  & 0.850  & 0.851  \\
                                           & r                         & 0.922  & 0.922  & 0.922  & 0.922  & 0.922  \\ \cline{2-7}
            \multirow{4}{*}{QIENet\_Conv6} & RMSE ($\mathrm{W/m^{2}}$) & 95.39  & 95.61  & 95.69  & 95.54  & 95.44  \\
                                           & MBE ($\mathrm{W/m^{2}}$)  & -0.16  & -0.86  & -1.20  & -1.83  & 0.26   \\
                                           & $\mathrm{R^{2}}$          & 0.844  & 0.843  & 0.843  & 0.843  & 0.844  \\
                                           & r                         & 0.919  & 0.918  & 0.918  & 0.918  & 0.919  \\ \cline{2-7}
            \multirow{4}{*}{ERA5}          & RMSE ($\mathrm{W/m^{2}}$) & 131.79 & 131.79 & 131.79 & 131.79 & 131.79 \\
                                           & MBE ($\mathrm{W/m^{2}}$)  & 9.88   & 9.88   & 9.88   & 9.88   & 9.88   \\
                                           & $\mathrm{R^{2}}$          & 0.702  & 0.702  & 0.702  & 0.702  & 0.702  \\
                                           & r                         & 0.845  & 0.845  & 0.845  & 0.845  & 0.845  \\ \cline{2-7}
            \multirow{4}{*}{JAXA}          & RMSE ($\mathrm{W/m^{2}}$) & 93.03  & 93.03  & 93.03  & 93.03  & 93.03  \\
                                           & MBE ($\mathrm{W/m^{2}}$)  & 19.56  & 19.56  & 19.56  & 19.56  & 19.56  \\
                                           & $\mathrm{R^{2}}$          & 0.852  & 0.852  & 0.852  & 0.852  & 0.852  \\
                                           & r                         & 0.932  & 0.932  & 0.932  & 0.932  & 0.932  \\ \cline{2-7}
            \multirow{4}{*}{NSRDB}         & RMSE ($\mathrm{W/m^{2}}$) & 116.51 & 116.51 & 116.51 & 116.51 & 116.51 \\
                                           & MBE ($\mathrm{W/m^{2}}$)  & 11.87  & 11.87  & 11.87  & 11.87  & 11.87  \\
                                           & $\mathrm{R^{2}}$          & 0.771  & 0.771  & 0.771  & 0.771  & 0.771  \\
                                           & r                         & 0.887  & 0.887  & 0.887  & 0.887  & 0.887  \\ \hline
        \end{tabular}
    }
    \label{table:QIENet_SIndex}
\end{table}

\begin{table}[H]
    \caption{The average values of evaluation indicators at each station using QIENet\_FC1, QIENet\_Conv1, JAXA, ERA5, and NSRDB on the test dataset.}
    \centering
    \resizebox{0.6\linewidth}{!}{
        \begin{tabular}{ccccc}
            \hline
            Model         & RMSE ($\mathrm{W/m^{2}}$) & MBE ($\mathrm{W/m^{2}}$) & $\mathrm{R^{2}}$ & r           \\ \hline
            QIENet\_FC1   & 97.00±13.50               & -1.21±12.02              & 0.822±0.046      & 0.912±0.019 \\
            QIENet\_Conv1 & 91.96±13.55               & -2.06±10.22              & 0.841±0.037      & 0.921±0.015 \\
            ERA5          & 129.83±19.77              & 9.50±12.83               & 0.684±0.075      & 0.841±0.032 \\
            JAXA          & 90.67±18.73               & 19.12±14.40              & 0.844±0.064      & 0.935±0.026 \\
            NSRDB         & 115.12±19.95              & 11.66±15.88              & 0.751±0.111      & 0.885±0.046 \\ \hline
        \end{tabular}
    }
    \label{table:QIENet_SIndex_everystation}
\end{table}

\begin{table}[H]
    \caption{Pearson correlation coefficients (PCCs) between the hourly GHI observations and remote sensing data of each satellite channel.}
    \centering
    \resizebox{0.45\linewidth}{!}{
        \begin{tabular}{cccccc}
            \hline
            Satellite channel &  & $\mathrm{PCC_{max}}$ & $\mathrm{PCC_{min}}$ & $\mathrm{PCC_{mean}}$ & $\mathrm{PCC_{median}}$ \\ \hline
            B01            &  & 0.181                & 0.138                & 0.155                 & 0.152                   \\
            B02            &  & 0.180                & 0.124                & 0.148                 & 0.144                   \\
            B03            &  & 0.180                & 0.099                & 0.137                 & 0.132                   \\
            B04            &  & 0.198                & 0.139                & 0.183                 & 0.183                   \\
            B05            &  & 0.218                & 0.159                & 0.196                 & 0.196                   \\
            B06            &  & 0.205                & 0.124                & 0.172                 & 0.171                   \\
            B07            &  & 0.543                & 0.444                & 0.509                 & 0.505                   \\
            B08            &  & 0.019                & 0.022                & 0.024                 & 0.025                   \\
            B09            &  & 0.046                & 0.042                & 0.050                 & 0.051                   \\
            B10            &  & 0.103                & 0.077                & 0.097                 & 0.098                   \\
            B11            &  & 0.370                & 0.205                & 0.297                 & 0.295                   \\
            B12            &  & 0.359                & 0.210                & 0.297                 & 0.295                   \\
            B13            &  & 0.366                & 0.205                & 0.294                 & 0.293                   \\
            B14            &  & 0.345                & 0.192                & 0.276                 & 0.275                   \\
            B15            &  & 0.307                & 0.174                & 0.247                 & 0.246                   \\
            B16            &  & 0.250                & 0.145                & 0.204                 & 0.204                   \\ \hline
        \end{tabular}
    }
    \label{table:band_pearson}
\end{table}

\bibliography{./refs/mybibfile_abbreviated}

\end{document}